\documentclass[showpacs,twocolumn,preprintnumbers,showpacs,showkeys,superscriptaddress,amsmath,amssymb,nofootinbib]{revtex4}
\usepackage{amsmath}
\usepackage{latexsym}
\usepackage{pstricks,pst-coil}
\usepackage{color}
\usepackage{latexsym}
\usepackage{amsmath}
\usepackage{amssymb}
\usepackage{eufrak}
\usepackage{euscript}
\usepackage{pstricks}
\usepackage{graphics}
\usepackage{graphicx}
\usepackage{picture}

\newcommand{\be}{\begin{equation}}
\newcommand{\ee}{\end{equation}}
\newcommand{\ba}{\begin{eqnarray}}
\newcommand{\ea}{\end{eqnarray}}

\def\uma{\rm 1\!\!\hskip 1 pt l}
\begin{document}

\markright{TeV- and MeV-physics out of an $SU_{L}(2)\times U_{R}(1)_{J} \times U(1)_{K}$ model}

\title{\Large{TeV- and MeV-physics out of an $SU_{L}(2)\times U_{R}(1)_{J} \times U(1)_{K}$ model}}

\author{M. J. Neves}\email{mariojr@ufrrj.br}
\affiliation{Departamento de F\'isica, Universidade Federal Rural do Rio de Janeiro,
BR 465-07, 23890-971, Serop\'edica, RJ, Brazil}
\author{J. A. Hela\"yel-Neto}\email{helayel@cbpf.br}
\affiliation{Centro Brasileiro de Pesquisas F\'isicas, Rua Dr. Xavier Sigaud
150, Urca,\\
Rio de Janeiro, Brazil, CEP 22290-180}

\date{\today}



\begin{abstract}
\noindent

The standard electroweak interaction is here re-assessed to accommodate two different situations
in Particle Physics. The first one is a $Z'$-model at the TeV-scale physics. The second one tackles the recent
discussion of a possible fifth force mediated by a $17$-MeV X-boson associated with an electron-positron
emission in the transition of an excited 8-Beryllium to its ground state. The anomaly-free model that provides these two
scenarios is based on an $SU_{L}(2) \times U_{R}(1)_{J} \times U(1)_{K}$-symmetry. It yields a new massive neutral boson,
an exotic massive neutral fermion, right-neutrinos and an additional neutral Higgs particle, which stems from a supplementary
Higgs field, introduced along with the usual Higgs doublet responsible for the electroweak breaking and the masses of $W^{\pm}$ and $Z^{0}$.
Yukawa interactions of the two scalars generate the masses of the Standard Model leptons, neutrinos and a new exotic fermion of the model.
The vacuum expectation values of the Higgses fix up two independent energy scales. One of them is the well-confirmed electroweak
scale, $246 \, \mbox{GeV}$, whereas the other one is set up by adopting an experimental estimate for the $Z'$-mass.

\end{abstract}


\maketitle

\pagestyle{myheadings}
\markright{TeV- and MeV-physics out of an $SU_{L}(2)\times U_{R}(1)_{J} \times U(1)_{K}$ model}
%
%
%
%
%
%
%

\section{Introduction}

The understanding of a new physics beyond the Standard Model (SM) has been a challenge for High Energy Physics
over the past thirty years. The results of the ATLAS and CMS Collaborations at LHC point to the masses peaks that
could be explained by the existence of new particles in a scenario of a possible fifth fundamental interaction.
The study of $pp$ collisions at a center-of-mass energy $\sqrt{s}= 8 \, \mbox{TeV}$ and $\sqrt{s}= 13 \, \mbox{TeV}$
suggests the presence of hypothetical heavy $W^{\, \prime}$-  and $Z^{\, \prime}$-bosons in the particle spectrum
\cite{Atlas,CMS2015,CMS20171,CMS20172}. These results can signal the emergence of a new Physics at the $\mbox{TeV}$-scale.
The structure of the $W'$- and $Z'$-model is based on the left/right symmetric
$SU_{L}(2)\times SU_{R}(2) \times U(1)_{B-L}$ gauge symmetry with an extra Higgs sector to account for the heavy bosons at the $\mbox{TeV}$-scale \cite{DobrescuPRL,DobrescuJHEP}. The pure $Z'$-model is based on the gauge $SU_{L}(2)\times U_{Y}(1) \times U(1)_{B-L}$, in which the $U(1)_{B-L}$-extra can connect the particles of the SM with a content of the Dark Matter \cite{MorettiarXiv2017,LangackerRMP2009,Kanemura2017}.


Recently, in a different experimental context, anomalies in the nuclear decay of the excited state of $8\, \mbox{Be}^{\ast}$ to its ground state
is suggesting the existence of a new neutral $X$-boson in the decay process $8 \, \mbox{Be}^{\ast} \rightarrow 8 \, \mbox{Be} + X$
\cite{KrasPRL2016}. The $X$-boson immediately decays into an electron-positron pair $X \rightarrow e^{+} \, + \, e^{-}$.
It exhibits a vector nature, like the electromagnetic (EM) photon, but its mass must be approximately $M_{X}=17 \, \mbox{MeV}$.
Other important property is that the $X$-boson mixes kinetically with the usual photon in the gauge sector of the model.
Furthermore, the vector current for fermions of the SM interacting with the $X$-boson has a coupling with weaker magnitude
for protons in relation to neutrons. These are called protophobic interactions. In principle, the $X$-boson has an Abelian nature,
so it can be described by an extra gauge group $U(1)_{X}$, that mixes with the EM gauge group $U(1)_{em}$. Therefore,
an Extended SM with the presence of the $X$-boson could be described by the unification of the gauge group
$SU_{c}(3) \times SU_{L}(2) \times U_{R}(1)_{Y} \times U(1)_{X}$. Nowadays, there is a huge number of alternative models that
describe this unification \cite{JFeng2016PRL,KitaharaPRD2017,GuHe2017NPB,JFeng2016PRD,RosearXiv2017,SetoarXiv2017}.

The appearance of such a new
boson mediating an interaction with typical range of $12$ fm may put into evidence a fifth fundamental interaction in Nature. On the other hand,
there has been in the literature a great deal of interest on the activity related to the phenomenology of the so-called hidden sector
para-photons and milli-charged particles \cite{Ahlers2007,Jaeckel2008,Arias2010}. In this rich framework, we pursue an investigation to understand
whether our proposal could fit to describe physics at the Sub-eV scale of the para-photons.

In this contribution, we built up a gauge model with an $SU_{L}(2) \times U_{R}(1)_{J} \times U(1)_{K}$ symmetry group that is twofold, according
to the spontaneous symmetry breaking (SSB) pattern, as it shall become clear in their sequel. The scalar sector consists of the usual electroweak
Higgs doublet and an extra $SU_{L}(2)$-singlet Higgs  : the extra Higgs may break the symmetry $U(1)_{K}$ above or below the $246 \, \mbox{GeV}$
electroweak scale. These two possibilities open up new scenarios that may accommodate the extra Higgs and its associated extra gauge boson in
different scales : $\mbox{MeV}$- , $\mbox{GeV}$- and $\mbox{TeV}$-scales, according to the choice of the VEV-scales parameters.
The first case, when the SSB is above the $246$ GeV-scale, setting up to accommodate the $Z'$-heavy boson as a candidate to the
particle at the TeV-scale. The second one, when the SSB is below the $246$ GeV, this is the scenario to accommodate the $X$-boson of
$17$ MeV, and also the para-photon hidden particle at the lower Sub-eV scale. The field content and the quantum numbers attributed to the
particles ensure gauge anomaly cancellation.



\section{The $\mbox{TeV}$-physics and the $Z'$-boson}

The fermion matter sector is described by the Lagrangian :
\begin{equation}\label{Lleptons}
{\cal L}_{f}=\bar{\Psi}_{L}\, i \, \, \slash{\!\!\!\!D} \, \Psi_{L}
+\bar{\Psi}_{R} \, i \, \, \slash{\!\!\!\!D} \, \Psi_{R}
+\bar{\nu}_{R} \, i \, \, \slash{\!\!\!\!D} \, \nu_{R}
+\bar{\zeta} \, i \, \, \slash{\!\!\!\!D} \, \zeta  \; ,
\end{equation}
where we have introduced the $\zeta$-fermion associated with the $U(1)_{K}$ group.
The $\Psi_{L}$ is defined as a left-handed doublet of fermions of the SM, neutrinos/leptons $L:=( \nu_{\ell L} \; \; \ell_{L} )^{t}$
or quarks-up/down $q_{L}:=( u_{L} \; \; d_{L} )^{t}$ ,
that turn in the fundamental representation of $SU_{L}(2)$, $\Psi_{R}=\{ \, \ell_{R} \, , \, u_{R} \, , \, d_{R} \, \}$,
$\nu_{\ell R}$ and $\zeta$ are singlets of Abelian sectors. The notation $\ell$
indicates the leptons of the SM, {\it i. e.}, $\ell=\left( \, e \, , \, \mu \, , \, \tau \, \right)$, and neutrinos family
$\nu_{\ell}=\left( \, \nu_{e} \, , \, \nu_{\mu} \, , \, \nu_{\tau} \, \right)$. The covariant derivatives acting on fermions are defined as given below
\begin{eqnarray}\label{DmuPsi}
D_{\mu}\Psi_{L} \!\! &=& \!\! \left(\partial_{\mu}+i \, g \, A_{\mu}^{\, a} \, \frac{\sigma^{a}}{2} +i \, J_{L} \, g^{\prime} \, B_{\mu}
+ i \, K_{L} \, g^{\prime \prime} \, C_{\mu} \! \right) \! \Psi_{L} \; ,
\nonumber \\
D_{\mu}\Psi_{R} \!\! &=& \!\! \left( \phantom{\frac{1}{2}} \!\!\!\! \partial_{\mu} + i \, J_{R} \, g^{\prime} \, B_{\mu}+i \, K_{R} \, g^{\prime \prime} \, C_{\mu} \right) \Psi_{R} \; ,
\nonumber \\
D_{\mu}\nu_{\ell R} \!\! &=& \!\! \left( \phantom{\frac{1}{2}} \!\!\!\! \partial_{\mu} + i \, J_{\nu_{R}} \, g^{\prime} \, B_{\mu}+i \, K_{\nu_{R}} \, g^{\prime \prime} \, C_{\mu} \right) \nu_{\ell R} \; ,
\nonumber \\
D_{\mu}\zeta\!\! &=& \!\! \left( \phantom{\frac{1}{2}} \!\!\!\! \partial_{\mu}+ i \, J_{\zeta} \, g^{\prime} \, B_{\mu} + i \, K_{\zeta} \, g^{\prime \prime} \, C_{\mu} \right) \zeta  \; ,
\end{eqnarray}
in which $A^{\mu \, a}=\left(A^{\mu \, 1} , A^{\mu \, 2} , A^{\mu \, 3}\right)$ are the gauge fields of $SU_{L}(2)$, $B^{\mu}$ is the Abelian gauge field of $U_{R}(1)_{J}$, and $C^{\mu}$ the similar one to $U(1)_{K}$. Here the symbol $J$ stands for the generators of $U_{R}(1)_{J}$, whereas
$K$ represents the generator of $U(1)_{K}$, and the Pauli matrices $\frac{\sigma^{a}}{2} \, (a=1,2,3)$. In (\ref{DmuPsi}), $g$, $g^{\, \prime}$ and
$g^{\, \prime\prime}$ are dimensionless coupling constants of $SU_{L}(2)$, $U_{R}(1)_{J}$ and $U(1)_{K}$, respectively.

In the sector gauge fields, the field strength tensors are defined by
\begin{eqnarray}\label{Fmunu}
F_{\mu\nu} \!&=&\! \partial_{\mu}A_{\nu}-\partial_{\nu}A_{\mu}+i \, g \, \left[ \, A_{\mu} \, , \, A_{\nu} \, \right]  \; ,
\nonumber \\
B_{\mu\nu} \!&=&\! \partial_{\mu}B_{\nu}-\partial_{\nu}B_{\mu} \; ,
\nonumber \\
C_{\mu\nu}  \!&=&\! \partial_{\mu}C_{\nu}-\partial_{\nu}C_{\mu} \; .
\end{eqnarray}
%
%
%
%
%
The sector of gauge bosons is described by the Lagrangian
\begin{equation}\label{Lgauge}
{\cal L}_{gauge}=-\frac{1}{2} \,\mbox{tr}\left(F_{\mu\nu}^{\; 2}\right)
-\frac{1}{4} \, B_{\mu\nu}^{\; 2}
-\frac{1}{4} \, C_{\mu\nu}^{\; 2} \; .
\end{equation}

The Higgs sector is governed by the Lagrangian
\begin{eqnarray}\label{LHiggs}
{\cal L}_{Higgs}\!\!&=&\!\!
\left(D_{\mu}\Xi\right)^{\dagger} D^{\mu} \Xi
-\mu_{\Xi}^{\, 2} \, |\Xi|^2-\lambda_{\Xi} \, |\Xi|^{4}
\nonumber \\
&&
\hspace{-0.5cm}
+\left(D_{\mu}\Phi\right)^{\dagger} D^{\mu} \Phi
-\mu_{\Phi}^{\, 2} \, |\Phi|^2\!-\!\lambda_{\Phi} \, |\Phi|^{4}
\!-\!\lambda \, |\Xi|^2 \, |\Phi|^2
\nonumber \\
&&
\hspace{-0.5cm}
- \, y_{\ell} \, \bar{L} \, \Phi \, \ell_{R}
- \, y_{\ell}^{\ast} \, \bar{\ell}_{R} \, \Phi^{\dagger} \, L
\nonumber \\
&&
\hspace{-0.5cm}
- \, x_{\nu_\ell} \, \bar{L} \, \widetilde{\Phi} \, \nu_{\ell R}
- \, x_{\nu_\ell}^{\ast} \, \bar{\nu}_{\ell R} \, \widetilde{\Phi}^{\dagger} \, L
\nonumber \\
&&
\hspace{-0.6cm}
- \, f^{(d)} \, \bar{q}_{L} \, \Phi \, d_{R}
- f^{(d)\ast} \bar{d}_{R} \, \Phi^{\dagger} q_{L}
\nonumber \\
&&
\hspace{-0.6cm}
- \, f^{(u)} \, \bar{q}_{L} \, \widetilde{\Phi} \, u_{R}
- f^{(u)\ast} \bar{u}_{R} \, \widetilde{\Phi}^{\dagger} q_{L}
\nonumber \\
&&
\hspace{-0.5cm}
- \, z_{\nu_\ell} \, \bar{L} \, \widetilde{\Phi} \, \zeta_{R}
- \, z_{\nu_\ell}^{\ast} \, \bar{\zeta}_{R} \, \widetilde{\Phi}^{\dagger} \, L
\nonumber \\
&&
\hspace{-0.5cm}
- \, w \, \bar{\zeta}_{L} \, \Xi \, \zeta_{R}
- \, w^{\ast} \, \bar{\zeta}_{R} \, \Xi^{\dagger} \, \zeta_{L}
\nonumber \\
&&
\hspace{-0.6cm}
- \, t_{\nu_\ell} \, \bar{\zeta}_{L} \, \Xi \, \nu_{\ell R}
- \, t_{\nu_\ell}^{\ast} \, \bar{\nu}_{\ell R} \, \Xi^{\dagger} \, \zeta_{L}
\; ,
\end{eqnarray}
where $\mu_{\Xi}$, $\mu_{\Phi}$, $\lambda_{\Xi}$, $\lambda_{\Phi}$ and $\lambda$ are real parameters,
$\{ \, x_{\nu_\ell} \, , \, y_{\ell} \, , \, z_{\nu_\ell} \, , \, w \, , \, t_{\nu_\ell} \, , \, f^{(d)} \, , \, f^{(u)} \, \}$ are
Yukawa complex coupling constants that yield the masses for the fermions of the model, and $\widetilde{\Phi}:=i \, \sigma_{2} \, \Phi^{\ast}$.
The gauge symmetry allows to introduce these Yukawa interactions that mix the Dirac neutrinos with the $\zeta$-fermion.


The covariant derivative of (\ref{LHiggs}) acts on the $\Xi$-Higgs
as follows :
\begin{eqnarray}\label{DmuPhi1}
D_{\mu} \Xi \!\!&=&\!\! \left( \phantom{\frac{1}{2}} \!\!\!\! \partial_{\mu}+i \, J_{\Xi} \, g^{\prime} \, B_{\mu}+i \, K_{\Xi} \, g^{\prime\prime}
\, C_{\mu}\right) \Xi \; ,
\nonumber \\
D_{\mu} \Phi \!\!&=&\!\! \left( \, \partial_{\mu}
+ i \, g \, A_{\mu}^{\, a} \, \frac{\sigma^{a}}{2}+ i \, g'  \, J_{\Phi} \, B_{\mu} \, \right) \Phi \; .
\end{eqnarray}
The $\Phi$-field is a doublet in the fundamental representation of $SU_{L}(2)$, while $\Xi$-field is a scalar singlet of the
$SU_{L}(2)$ group, but the two scalars are charged under the Abelian sectors.
The Yukawa interactions are gauge invariant if we impose the relations :
\begin{eqnarray}
U_{R}(1)_{J}
:
\left\{
\begin{array}{ll}
-J_{\ell_{L}}+J_{\Phi}+J_{\ell_{R}}=0 \\
-J_{q_{L}}+J_{\Phi}+J_{d_{R}}=0 \\
-J_{q_{L}}-J_{\Phi}+J_{u_{R}}=0 \\
-J_{\zeta_{L}}+J_{\Xi}+J_{\zeta_{R}}=0 \\
-J_{\zeta_{L}}+J_{\Xi}+J_{\nu_{R}}=0
\end{array}
\right.
\; ,
\end{eqnarray}
\begin{eqnarray}
U(1)_{K}
: \left\{
\begin{array}{ll}
-K_{\ell_{L}}+K_{\Phi}+K_{\ell_{R}}=0 \\
-K_{q_{L}}+K_{\Phi}+K_{d_{R}}=0 \\
-K_{q_{L}}-K_{\Phi}+K_{u_{R}}=0 \\
-K_{\zeta_{L}}+K_{\Xi}+K_{\zeta_{R}}=0 \\
-K_{\zeta_{L}}+K_{\Xi}+K_{\nu_{R}}=0
\end{array}
\right.
\; .
\end{eqnarray}

The minimal value of the Higgs potential at the $(|\Xi|,|\Phi|)$-plane is obtained by the non-trivial VEVs. The $\Xi$-Higgs VEV
is defined by $\langle \Xi \rangle_{0}=u/\sqrt{2}$, and for the $\Phi$-Higgs, the VEV-scale is chosen as
$ \langle \Phi \rangle_{0}=
\left(
\begin{array}{c}
0 \\
\frac{v}{\sqrt{2}} \\
\end{array}
\right)$,
where $u$- and $v$-scales are given by
\begin{eqnarray}
u \!&\simeq&\! \sqrt{-\frac{\mu_{\Xi}^{2}}{\lambda_{\Xi}}} \left( 1- \frac{\lambda}{4 \, \lambda_{\Phi}}
\frac{\mu_{\Phi}^{2}}{\mu_{\Xi}^{2}} \right) \; ,
\nonumber \\
v \!&\simeq&\! \sqrt{-\frac{\mu_{\Phi}^{2}}{\lambda_{\Phi}}} \left( 1- \frac{\lambda}{4 \, \lambda_{\Xi}}
\frac{\mu_{\Xi}^{2}}{\mu_{\Phi}^{2}} \right)  \; ,
\end{eqnarray}
when $\mu_{\Xi}^{2}<0$ and $\mu_{\Phi}^{2}<0$. Here we have assumed that the coupling between the scalar fields is weaker in relation
to others one, {\it i. e.}, $\lambda \ll (\lambda_{\Xi},\lambda_{\Phi})$. Thus the VEV-scales have a correction due to
mixed $\lambda$-coupling of the $\Xi$- and $\Phi$-fields. Clearly, when $\lambda \rightarrow 0$, we obtain
$u=\sqrt{-\frac{\mu_{\Xi}^{2}}{\lambda_{\Xi}}}$ and $v=\sqrt{-\frac{\mu_{\Phi}^{2}}{\lambda_{\Phi}}}$.

We adopt the parametrization of the
$\Xi$- and $\Phi$-complex fields in the unitary gauge as below
\begin{eqnarray}\label{PhiGaugeparametrization}
\Xi(x) = \frac{u+\tilde{F}(x)}{\sqrt{2}} \; , \;
\Phi(x) = \frac{v+\tilde{H}(x)}{\sqrt{2}} \,
\left(
\begin{array}{c}
0 \\
1 \\
\end{array}
\right) \; ,
\end{eqnarray}
where $\tilde{F}$ and $\tilde{H}$ are real functions.

The VEVs-$\{ \, u \, , \, v \, \}$ define two independent scales for the breaking of the gauge symmetry.
In this scenario, we perform the SSB following the sequence
\begin{equation}
SU_{L}(2) \times U_{R}(1)_{J} \times U(1)_{K} \! \stackrel{\langle \Xi \rangle}{\longmapsto}\!
SU_{L}(2) \times U_{Y}(1) \stackrel{\langle \Phi \rangle}{\longmapsto} U_{em}(1) \; ,
\end{equation}
such that the condition $u \gg v$ must be satisfied. Therefore, we can associate the $u$-VEV with the TeV-scale, and the $v$-VEV
is the usual scale of the SM, {\it i. e.}, $v=246 \, \mbox{GeV}$. Thus, after the SSB, we get the gauge sector Lagrangian
\begin{eqnarray}\label{LGaugemassesXB}
{\cal L}_{gauge} \!\!&=&\!\!
-\frac{1}{2} \, W_{\mu\nu}^{+} \, W^{\mu\nu \, -} +m_{W}^{\, 2} \, W_{\mu}^{\;+}W^{\mu-}
\nonumber \\
&&
\hspace{-1cm}
-\frac{1}{4}\left(\partial_{\mu}A_{\nu}^{\,3}-\partial_{\nu}A_{\mu}^{\,3} \right)^{2}
-\frac{1}{4} \, B_{\mu\nu}^{\, 2}
-\frac{1}{4} \, C_{\mu\nu}^{\, 2}
\nonumber \\
&&
\hspace{-1cm}
+\frac{1}{2} \, \frac{v^{2}}{4} \left( \phantom{\frac{1}{2}} \!\!\!\!\! g^{\prime} \, B_{\mu}-g \, A_{\mu}^{3} \right)^{2}
\!\!+\frac{u^2}{2} \left( \phantom{\frac{1}{2}} \!\!\!\! g^{\prime} \, B_{\mu}- \, g^{\prime \prime} \, C_{\mu} \right)^2 \! ,
\hspace{0.8cm}
\end{eqnarray}
where, for convenience, we have chosen that $J_{\Xi}+K_{\Xi}=0$ to keep the $\Xi$-Higgs uncharged.
This sector introduces the $SO(2)$- transformations in (\ref{LGaugemassesXB}) to eliminate the mixed terms
\begin{eqnarray}\label{transfA0CGY}
B_{\mu} \!&=&\! \cos\alpha \, \tilde{Z}'_{\mu}+ \sin\alpha \, Y_{\mu}
\nonumber \\
C_{\mu} \!&=&\! -\sin\alpha \, \tilde{Z}'_{\mu}+\cos\alpha \, Y_{\mu} \; ,
\nonumber \\
A_{\mu}^{3}&=& \cos\theta_{W} \, \tilde{Z}_{\mu} + \sin\theta_{W} \, A_{\mu}
\nonumber \\
Y_{\mu}&=&-\sin\theta_{W} \, \tilde{Z}_{\mu} + \cos\theta_{W} \, A_{\mu} \; ,
\end{eqnarray}
where $\alpha$ is a mixing angle between $U(1)'s$ gauge fields, it is parameterized by the relation
\begin{eqnarray}
g_{Y}=g^{\, \prime} \sin\alpha=g^{\, \prime\prime} \cos\alpha \; .
\end{eqnarray}
In this stage, $g_{Y}$ is the coupling constant of $U_{Y}(1)$, after the first SSB. The hypercharge generator satisfies
the relation $Y=J+K$, in which $Y$ is constituted by the primitive charge generators $J$ and $K$ of the model.
The other $\theta_{W}$-mixing angle is the Weinberg's angle, that satisfies the relation
\begin{eqnarray}\label{eg1g}
e = g \, \sin\theta_{W}=g_{Y} \, \cos\theta_{W} \; .
\end{eqnarray}
%
%
Naturally, the electric charge satisfies the relation $Q_{em}=I_{3}+Y$, where $I_{3}=\frac{\sigma_{3}}{2}$. For example,
the $\Xi$-Higgs is a singlet of $SU_{L}(2)$, thus $I^{3}=0$ and it hypercharge is $Y_{\Xi}=J_{\Xi}+K_{\Xi}=0$, so the electric charge is $Q_{em}^{\,(\Xi)}=0$. The model is quiral anomaly free in accord with the charges
displayed in the table \ref{Table1}.
\begin{table}
\begin{tabular}{|l|l|l|l|l|l|}
\hline
\mbox{Fields} \& \mbox{particles} & $Q_{em}$ & $I^{3}$ & $Y$ & $J$ & $K$ \\
\hline
\mbox{lepton-left} & $-1$ & $-1/2$ & $-1/2$ & $ 0 $ & $-1/2 $  \\
\hline
\mbox{neutrino-left} & $0$ & $+1/2$ & $-1/2$ & $ 0 $ & $ -1/2 $ \\
\hline
\mbox{lepton-right} & $-1$ & $0$ & $-1$ & $-1/2$ & $ -1/2 $ \\
\hline
\mbox{neutrino-right} & $0$ & $0$ & $0$ & $+1/2$ & $-1/2$ \\
\hline
$\zeta$-\mbox{left} & $0$ & $0$ & $0$ & $-1/2$ & $+1/2$ \\
\hline
$\zeta$-\mbox{right} & $0$ & $0$ & $0$ & $+1/2$ & $-1/2$ \\
\hline
\mbox{u-quark-left} & $+2/3$ & $+1/2$ & $+1/6$ & $0$ & $+1/6$  \\
\hline
\mbox{d-quark-left} & $-1/3$ & $-1/2$ & $+1/6$ & $0$ & $+1/6$ \\
\hline
\mbox{u-quark-right} & $+2/3$ & $0$ & $+2/3$ & $+1/2$ & $+1/6$ \\
\hline
\mbox{d-quark-right} & $-1/3$ & $0$ & $-1/3$ & $-1/2$ & $+1/6$ \\
\hline
$W^{\pm}$-\mbox{bosons} & $\pm \, 1$ & $\pm \, 1$ & $0$ & $0$ & $0$ \\
\hline
\mbox{neutral bosons} & $0$ & $0$ & $0$ & $0$ & $0$  \\
\hline
$\Xi$-\mbox{Higgs} & $0$ & $0$ & $0$ & $-1$ & $+1$ \\
\hline
$\Phi$-\mbox{Higgs} & $0$ & $-1/2$ & $+1/2$ & $+1/2$ & $0$ \\
\hline
\end{tabular}
\caption{The particle content for the $Z'$-model candidate at the TeV-scale physics.
The $J$- and $K$-charges are such that anomalies cancel out.}\label{Table1}
\end{table}

As in the usual case, the mass of $W^{\pm}$ is $m_{W}= g \, v/2$, and mass of $\tilde{Z}'$
is identified in terms of VEV scale-$u$ as $m_{\tilde{Z}'}=g^{\prime} \, u/\cos\alpha$. The $\tilde{Z}$- and $\tilde{Z}'$- gauge sector
is so mixed by the mass matrix in (\ref{LGaugemassesXB})
%
%
%
\begin{eqnarray}\label{MZX}
M_{\tilde{Z}-\tilde{Z}'}^{2}=\left(
\begin{array}{cc}
m_{\tilde{Z}}^{\, 2} & -\frac{m_{\tilde{Z}} \, m_{\tilde{Z}'}}{4x}
\\
\\
-\frac{m_{\tilde{Z}} \, m_{\tilde{Z}'}}{4x} & m_{\tilde{Z}'}^{\, 2}  \\
\end{array}
\right) \; .
\end{eqnarray}
Here, the dimensionless parameter $x$ has been defined as $x:=u/v\,\cos^2\alpha$,
and $m_{\tilde{Z}}$ is the mass of $\tilde{Z}$ at the tree-level approximation
$m_{W}=m_{\tilde{Z}}\cos\theta_{W}$.  The VEV-scale $v$ is defined by the Fermi's constant
by $v=\left(\sqrt{2} \, G_{F}\right)^{-1/2}\simeq 246 \, \, \mbox{GeV}$,
and considering that Weinberg's angle has the experimental value $\sin^{2}\theta_{W} \simeq 0.23$,
the parametrization (\ref{eg1g}) gives us the masses of $W^{\pm}$, $\tilde{Z}$ and $\tilde{Z}'$ in terms
of the fundamental constants. Thus, the $W$-mass is given like in the electroweak model
\begin{eqnarray}\label{massW}
m_{W} \!\!&=&\!\! \frac{37 \, \, \mbox{GeV} }{|\sin\theta_{W}|} \simeq 77 \, \, \mbox{GeV} \; .
\end{eqnarray}
In the mixed neutral sector, $\tilde{Z}$- and $\tilde{Z}'$-fields are not physical fields of $Z^{0}$ and $Z'$ yet.
In fact, the real physical fields come from the orthogonal transformation that mixes $\tilde{Z}$- and $\tilde{Z}'$-fields
\begin{eqnarray}
\left(
\begin{array}{c}
\tilde{Z}'_{\mu} \\
\tilde{Z}_{\mu} \\
\end{array}
\right)
=\left(
\begin{array}{cc}
\cos\theta_{ZZ'} & \sin\theta_{ZZ'}
\\
\\
-\sin\theta_{ZZ'} & \cos\theta_{ZZ'} \\
\end{array}
\right)
\left(
\begin{array}{c}
Z'_{\mu} \\
Z_{\mu} \\
\end{array}
\right) \; ,
\end{eqnarray}
where $\theta_{ZZ'}$ is the mixing angle defined by
\begin{eqnarray}\label{thetaZZm}
\tan2\theta_{ZZ'}=-\frac{1}{2x} \, \frac{m_{\tilde{Z}} \, m_{\tilde{Z'}}}{m_{\tilde{Z'}}^2-m_{\tilde{Z}}^2} \; .
\end{eqnarray}
Thereby, the $m_{\tilde{Z}-\tilde{Z}'}^{(\pm)}$-eigenvalues of the mass matrix (\ref{MZX}) are given by
\begin{eqnarray}
m_{\tilde{Z}-\tilde{Z}'}^{(\pm) \, 2}= \frac{m_{\tilde{Z}}^{2}+m_{\tilde{Z}'}^{2}\pm\sqrt{(m_{\tilde{Z}}^{2}-m_{\tilde{Z}'}^{2})^{2}
+\frac{m_{\tilde{Z}}^{2} \, m_{\tilde{Z}'}^{2}}{4 x^2}}}{2} \, .
\hspace{0.3cm}
\end{eqnarray}
Explicitly, the masses of $Z$- and $Z'$-bosons with the correction on the VEV-scales ratio $v/u$, and for the condition
$m_{\tilde{Z}'} \gg m_{\tilde{Z}}$, we get the results
\begin{eqnarray}\label{MassZX}
M_{Z^{0}}=m_{Z-Z'}^{(-)} \!\!& \simeq &\!\! \frac{e \, v}{\sin2\theta_{W}} \! \left( 1 - \frac{1}{32 x^2} \right) \; ,
\nonumber \\
M_{Z'}=m_{Z-Z'}^{(+)} \!\!& \simeq &\!\! m_{\tilde{Z}'} \simeq \frac{2 \, e \, u}{\sin2\alpha \cos\theta_{W}} \; .
\hspace{0.8cm}
\end{eqnarray}
The $m_{Z-Z'}^{(+)}$-eigenvalue is the $Z^{0}$-mass with the correction of $v/u$. Using the previous values for the parameters $e$, $v$ and
$\theta_{W}$, the theoretical $Z$-mass is around the
\begin{eqnarray}\label{MassZ}
M_{Z^{0}} \simeq 89 \, \left( 1 - \frac{v^{2}}{32 \, u^2} \, \cos^{4}\alpha \right) \mbox{GeV} \, .
\end{eqnarray}

We know the experimental value of $Z^{0}$-mass is $M_{Z^{0}}^{(exp)}=91.1876 \, \pm \, 0.0021 \;  \mbox{GeV}$, in which the theoretical values in (\ref{massW}) and (\ref{MassZ}) were obtained at the tree level approximation. The actual experimental values include radiative corrections, which we are not computing explicitly here. The radiative corrections give the $\theta_{W}$-value around the $\sin^2\theta_{W}\simeq0.21$, and the fine
structure constant as function of $W$-mass is $\alpha^{-1}(m_{W})=127.49$. Thereby, the theoretical $W$- and $Z$-masses
are in accord with the experimental data by the results
\begin{eqnarray}\label{massWZTeo}
m_{W} \!&\simeq&\! \frac{38 \, \, \mbox{GeV} }{|\sin\theta_{W}|} \simeq 83 \pm 2.4 \, \mbox{GeV} \; .
\nonumber \\
m_{Z} \!&=&\! \frac{m_{W}}{\cos\theta_{W}} \simeq 93.8 \pm 2.0 \, \mbox{GeV} \; .
\end{eqnarray}

The parameters $u$-scale and $\alpha$-angle are undeterminate in the previous $Z$- and $Z'$-masses.
Thereby, the ratio between the masses from (\ref{MassZX}) is given by
\begin{eqnarray}
\frac{M_{Z'}}{M_{Z^{0}}} \simeq \frac{u}{\sin(2\alpha)} \, \frac{4\sin\theta_{W}}{v} \simeq \frac{u}{\sin(2\alpha)} \, 0.007 \, \mbox{GeV}^{-1} .
\end{eqnarray}
The recent papers of the CMS Collaboration points to the hypothetical $Z'$ have upper limits
that excludes at $95\%$ confidence level masses below the $2.0 \, \mbox{TeV}$ \cite{CMS20172}.
Therefore, we fix the $Z$- and $Z'$-masses as $M_{Z'}=2.0 \, \mbox{TeV}$ and $M_{Z^{0}}=91 \, \mbox{GeV}$
to estimate the ratio of the $u$-scale by the $\alpha$-angle, {\it i. e.}, $u \simeq 2.8 \times \sin(2\alpha) \, \mbox{TeV}$.
Thereby, the maximum value for the VEV-scale is $u=2.8 \, \mbox{TeV}$, when $\alpha=45^{o}$.


%
Under these conditions, the $\theta_{ZZ'}$-mixing angle in (\ref{thetaZZm}) is estimated by
\begin{eqnarray}
\tan 2 \theta_{ZZ'}\simeq -\frac{m_{\tilde{Z}}}{2 \, m_{\tilde{Z'}}} \frac{v}{u} \cos^2\alpha \simeq - \, 10^{-3} \; ,
\end{eqnarray}
{\it i. e.}, $\theta_{ZZ'}\simeq -0.028$.

%
%
%

\begin{figure}[h]
\centering
\includegraphics[scale=0.4]{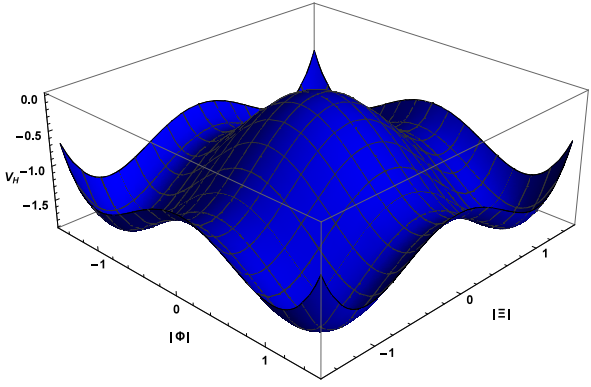}
\caption{The Higgs potential as function of the variables $|\Xi|$ and $|\Phi|$. We use the estimated values : $\mu_{\Xi}^2\simeq-0.8 \, \mbox{TeV}^{2}$, 
and when $\mu_{\Phi}^2\simeq-0.8 \, \mbox{TeV}^{2}$, $\lambda_{\Phi}\simeq 0.5 $ and $\lambda\simeq 10^{-5} $. The degenerated vacuum of the Higgs fields are illustrated by the four down peaks.} \label{FIGHiggsPotential}
\end{figure}

The sector of the Higgs fields $\tilde{F}-\tilde{H}$ after the SSBs is the following mass matrix comes out :
%
%
%
\begin{eqnarray}\label{MFH}
M_{\tilde{F}-\tilde{H}}^{\, 2}=\!
\left( \!\!
\begin{array}{cc}
m_{\tilde{F}}^{\, 2}  & - \, \frac{\tilde{\lambda}}{2} \, m_{\tilde{H}} \, m_{\tilde{F}}
\\
\\
-\, \frac{\tilde{\lambda}}{2} \, m_{\tilde{H}} \, m_{\tilde{F}} & m_{\tilde{H}}^{\, 2}   \\
\end{array}
\!\!\right) . \;\;
\end{eqnarray}
Here, $m_{\tilde{F}}=\sqrt{2 \lambda_{\Xi} \, u^{2}}$ and $m_{\tilde{H}}=\sqrt{2 \lambda_{\Phi} \, v^{2}}$ are the masses of $\tilde{F}$- and $\tilde{H}$-fields, when $\lambda \rightarrow 0$, and $\tilde{\lambda}:=\lambda/\sqrt{\lambda_{\Phi}\lambda_{\Xi}}$ for short.
The $\tilde{F}$- and $\tilde{H}$-scalars are not the physical fields due to the non-diagonal mass matrix (\ref{MFH}). In fact,
the scalars physical fields are defined by the orthogonal transformations
\begin{eqnarray}
\left(
\begin{array}{c}
\tilde{F} \\
\tilde{H} \\
\end{array}
\right)
=\left(
\begin{array}{cc}
\cos\vartheta & \sin\vartheta
\\
\\
-\sin\vartheta & \cos\vartheta \\
\end{array}
\right)
\left(
\begin{array}{c}
F \\
H \\
\end{array}
\right) \; ,
\end{eqnarray}
where $\vartheta$ is the mixing angle defined by
\begin{eqnarray}\label{thetaZZm}
\tan2\vartheta=-\frac{\tilde{\lambda} \, m_{\tilde{H}} \, m_{\tilde{F}}}{m_{\tilde{F}}^{\, 2}-m_{\tilde{H}}^{\, 2}}\simeq
- \tilde{\lambda} \; \frac{m_{\tilde{H}}}{m_{\tilde{F}}} \; .
\end{eqnarray}
Thereby, the mass matrix (\ref{MFH}) yields the eigenvalues
\begin{equation}\label{autovaloresMHF}
m_{\tilde{F}-\tilde{H}}^{(\pm)\, 2}= \frac{ m_{\tilde{H}}^2  + m_{\tilde{F}}^2 \pm \sqrt{(m_{\tilde{F}}^2-m_{\tilde{H}}^2)^2+\tilde{\lambda}^2 \, m_{\tilde{H}}^2 \, m_{\tilde{F}}^2}}{2} \; .
\end{equation}
Whenever $m_{\tilde{F}} \gg m_{\tilde{H}}$, the $m_{\tilde{F}-\tilde{H}}^{(\pm)}$-eigenvalues of (\ref{MFH}) are
\begin{eqnarray}\label{MH}
M_{H}=m_{\tilde{F}-\tilde{H}}^{(-)} \!\!&\simeq&\!\! \sqrt{ 2 \, \lambda_{\Phi} \, v^{2}}
\, \left( 1-\frac{\lambda^2}{8 \, \lambda_{\Xi}\lambda_{\Phi}} \right) \; ,
\nonumber \\
M_{F}=m_{\tilde{F}-\tilde{H}}^{(+)} \!\!&\simeq&\!\! \sqrt{ 2 \, \lambda_{\Xi} \, u^{2}}
\, \left( 1+\frac{\lambda^2}{8 \, \lambda_{\Xi}^2} \, \frac{v^{2}}{u^{2}}  \right) \; .
\end{eqnarray}
The $m_{\tilde{F}-\tilde{H}}^{(-)}$ - eigenvalue is the Higgs mass of the SM with the correction that we consider
$\frac{\lambda^2}{8 \, \lambda_{\Xi} \, \lambda_{\Phi}} \ll 1$, then the $\lambda$-coupling constant
must have a weaker magnitude with respect to the others one. The experimental result for the $\Phi$-Higgs mass
$M_{H}= 125.7 \pm 0.4 \, \mbox{GeV}$ fixes the upper bound $\lambda/\lambda_{\Xi} \, \lambda_{\Phi} \lesssim 10^{-4}$.
Under these conditions, the Higgs potential from (\ref{LHiggs}) is depicted in the figure below (\ref{FIGHiggsPotential}).
%
%


If we adopt that $0.1 < \lambda_{\Xi} < 0.9$, the mass of $F$-Higgs is in the range
\begin{eqnarray}\label{MassH1u8GEV}
1.24 \, \mbox{TeV} < M_{F} < 3.7 \, \mbox{TeV} \; .
\end{eqnarray}
%
%
%
%
%

%
The sector of the leptons, neutrinos and $\zeta$-fermion acquires mass terms thanks to the Yukawa interactions.
After the SSB, the fermion sector can be cast into the form
\begin{eqnarray}\label{LpsichiM}
{\cal L}_{\ell-\nu_{\ell}-\zeta}=\bar{\ell} \, \left( i \, \slash{\!\!\!\partial} -m_{\ell} \, {\uma} \right) \, \ell +\bar{\xi} \, \left( i \, \slash{\!\!\!\partial} -M_{\nu_{\ell}-\zeta} \right) \, \xi \; ,
\end{eqnarray}
in which the lepton's mass is identified as $m_{\ell}\!=\!|y_{\ell}| \, v/\sqrt{2}$. The $\xi$-spinor is formed by $\xi^{\, t}=\left( \, \nu_{\ell} \, \, \, \zeta \, \right)^{t}$ and $M_{\nu_{\ell}-\zeta}$ is the mass matrix
\begin{equation}\label{Lleptonsmatrix}
M_{\nu_{\ell}-\zeta} =
\left(
\begin{array}{cc}
m_{\nu_\ell} \, {\uma} & \frac{|z_{\nu_{\ell}}|vR+|t_{\nu_{\ell}}|uL }{\sqrt{2}}
\\
\\
\frac{|z_{\nu_{\ell}}|vL+|t_{\nu_{\ell}}|uR }{\sqrt{2}} & m_{\zeta} \, {\uma}
\end{array}
\right) \; ,
\end{equation}
where $m_{\nu_{\ell}}\!=\!|x_{\nu_{\ell}}| \, v/\sqrt{2}$ and $m_{\zeta}\!=\!|w| \, u/\sqrt{2}$ are the masses of the $\zeta$-fermion and neutrinos,
when we make $|z_{\nu_{\ell}}|=|t_{\nu_{\ell}}| \rightarrow 0$.
Here, we write the constant couplings in terms of global phases that can absorbed through field reshuffling,
and $R$ and $L$ are the right- and left- projectors, that satisfy the relations $RL=LR=0$, $L^{2}=L$, $R^{2}=R$ and $L+R={\uma}$.
The neutrino/$\zeta$-fermion matrix can be diagonalized by a unitary transformation so that the masses are given by the $m_{\nu_{\ell}-\zeta}^{(\pm)}$-eigenvalues of (\ref{Lleptonsmatrix}) :
%
\begin{eqnarray}\label{mlzeta}
m_{\nu_{\ell}-\zeta}^{(\pm)}= \frac{ m_{\nu_\ell}+m_{\zeta} \!\pm\! \sqrt{\left(m_{\nu_\ell}-m_{\zeta} \right)^{2}+4 m_{\nu_\ell} m_{\zeta} |g_{f}|} }{2} \; ,
\hspace{0.3cm}
\end{eqnarray}
where
%
%
and $|g_{f}|$ is defined by the Yukawa coupling constants
\begin{eqnarray}
|g_{f}|=\frac{1}{|w|} \sum_{\ell\,= \, e \, , \, \mu \, , \, \tau} \frac{|t_{\nu_\ell}||z_{\nu_\ell}|}{|x_{\nu_\ell}|} \; .
\end{eqnarray}
Here we are in the scenario in that $u \gg v$, so the $\zeta$-fermion should be a heavier particle than any neutrino of the SM,
{\it i. e.}, $m_{\zeta} \gg  m_{\nu_\ell}$, and we consider also $|g_{f}| \ll 1$. Under these conditions, the eigenvalues
of (\ref{mlzeta}) are given by
\begin{eqnarray}\label{autoveloresmfermions}
M_{\nu_{\ell}}=m_{\ell-\zeta}^{(-)} \!&\simeq&\! \frac{|x_{\nu_\ell}| \, v}{\sqrt{2}} \left( \, 1+ |g_{f}| \, \right)
\nonumber \\
M_{\zeta}=m_{\nu_{\ell}-\zeta}^{(+)} \!&\simeq&\! \frac{|w| \, u}{\sqrt{2}} \left( \, 1+ |g_{f}| \,  \frac{m_{\nu_{\ell}}}{m_{\zeta}} \, \right)
\simeq \frac{|w| \, u}{\sqrt{2}}  \, .
\hspace{0.6cm}
\end{eqnarray}
The $m_{\nu_{\ell}-\zeta}^{(-)}$-eigenvalue gives us the neutrinos masses with the correction of the $|g_{f}|$-coupling constant.
The neutrinos masses to the squared are constraints by the subtraction between itself. For example, for the case of the
electron- and muon-neutrinos, this subtraction is
$|\Delta M_{\nu_{e}-\nu_{\mu}}^{2}|:=|M_{\nu_{e}}^2-M_{\nu_{\mu}}^2|\simeq \left( \, 7.53 \, \pm \, 0.18 \, \right) \times 10^{-5} \, \mbox{eV}^{2}$ \cite{ArakiPRL2005}.
Thus, the coupling constants are extremely weak $|\Delta x_{\nu_{e}-\nu_{\mu}}^{2}|:=||x_{\nu_{e}}|^2-|x_{\nu_{\mu}}|^2|\simeq 2.5 \times 10^{-27}$.

%
%

We point out that the $\zeta$-fermion introduced in our particle content yields, upon symmetry breaking and the fermion mass
matrix diagonalization, a neutral massive fermion with mass of the order of $550 \, \mbox{GeV}$ and no electromagnetic interaction.
It could therefore be considered an exotic fermion which is part of Dark Matter. The motivation to introduce the new $\zeta$-fermion
is the possible detection constraints of the Dark Matter with
vectors, scalars and axial-vectors mediators \cite{CMS20172}. For an axial-vector mediator like the $Z'$,
the Dark Matter constraint fixes a mass around less than $550 \, \mbox{GeV}$. Then, if we take a $\zeta$-mass of the order
$M_{\zeta}= 550 \, \mbox{GeV}$, we obtain $|w|\simeq 0.27$. Therefore, the $\nu-\zeta$-mixed Yukawa coupling constants are estimated as
%
%
\begin{eqnarray}
|\Delta t_{\nu_{e}-\nu_{\mu}}^{2}| \simeq |\Delta z_{\nu_{e}-\nu_{\mu}}^{2}| \lesssim 3.2 \times 10^{-17} \; .
\end{eqnarray}

The interactions of any fermion $\Psi$ with the neutral gauge bosons of the model are setting by
\begin{equation}\label{LintAZC}
{\cal L}^{\, int}= - \, e\, Q_{em} \, \bar{\Psi} \, \, \slash{\!\!\!\!A} \, \Psi
-e\, Q_{Z} \, \bar{\Psi} \, \, \slash{\!\!\!\!Z} \, \Psi
-e\, Q_{Z'} \, \bar{\Psi} \,\, \slash{\!\!\!\!Z'} \, \Psi \; .
\end{equation}
%
%
For convenience, we also define the $Q_{Z}$- and $Q_{Z'}$-effective charges associated with the interaction between fermions with $Z$- and $Z'$-bosons
\begin{eqnarray}
Q_{Z} \!\!&=&\!\! Q_{\tilde{Z}} \, \cos\theta_{ZZ'} + Q_{\tilde{Z'}} \, \sin\theta_{ZZ'}
\nonumber \\
Q_{Z'} \!\!&=&\!\! -Q_{\tilde{Z}} \, \sin\theta_{ZZ'} + Q_{\tilde{Z'}} \, \cos\theta_{ZZ'} \; ,
\end{eqnarray}
where $Q_{\tilde{Z}}$ and $Q_{\tilde{Z'}}$ are given by
\begin{eqnarray}\label{QZQZ'}
Q_{\tilde{Z}}\!\!&:=&\!\!\frac{1}{\sin\theta_{W}\cos\theta_{W}} \left( \phantom{\frac{1}{2}} \!\!\!\! I^{3}- Q_{em} \sin^{2}\theta_{W} \right) \; ,
\nonumber \\
Q_{\tilde{Z'}}\!\!&:=&\!\!\frac{1}{\sin2\alpha \cos\theta_{W}} \left( \phantom{\frac{1}{2}} \!\!\!\!\!\! - J \, + Y \sin^{2}\alpha \right) \; .
\end{eqnarray}
Using the charges from the table \ref{Table1} , we list below the interactions of $Z'$-boson with leptons, neutrinos and
$\zeta$-fermion :
\begin{eqnarray}\label{LllZ'}
{\cal L}_{\bar{\ell}\ell-Z'}^{\, int} \!=\! - \frac{e}{4} \frac{1-3\sin^2\alpha}{\sin2\alpha\cos\theta_{W}} \bar{\ell}
\,\, \slash{\!\!\!\!Z'} \ell
- \frac{e}{8} \, \frac{\cot\alpha}{\cos\theta_{W}} \, \bar{\ell} \,\, \slash{\!\!\!\!Z'} \gamma_{5}
\ell  \, ,
\hspace{0.3cm}
\end{eqnarray}
\begin{equation}\label{LnunuZ'}
{\cal L}_{\bar{\nu}_{\ell}\nu_{\ell}-Z'}^{\, int} \!=\! \frac{e}{8} \, \frac{1+\sin^2\alpha}{\sin\alpha\cos\theta_{W}} \, \bar{\nu}_{\ell} \, \, \slash{\!\!\!\!Z'} \nu_{\ell}
+\frac{e}{8} \, \frac{\cot\alpha}{\cos\theta_{W}} \, \bar{\nu}_{\ell} \, \, \slash{\!\!\!\!Z'} \gamma_{5} \nu_{\ell} \, ,
\end{equation}
\begin{eqnarray}\label{LzetaZ'}
{\cal L}_{\bar{\zeta}\zeta-Z'}^{\, int}\!=\! \frac{e}{2 \sin2\alpha \cos\theta_{W}} \, \bar{\zeta} \, \, \slash{\!\!\!\!Z'} \, \gamma_{5} \, \zeta
\, .
\end{eqnarray}

The recent $Z'$-phenomenology points to the cascade effects at the tree level using the CMS
data for the pp-collision at $\sqrt{s}=13 \, \mbox{TeV}$. The processes of the $Z'$-decay can be
useful to search the Dark Matter through the mono-V jets channels associated with the electroweak bosons $W$ or $Z$. The
observation of these final states could be interpreted as a Dark Matter particle content. The diagram for this effect
is illustrated below :
%
%
%
\begin{figure}[!h]
\begin{center}
\newpsobject{showgrid}{psgrid}{subgriddiv=1,griddots=10,gridlabels=6pt}
\begin{pspicture}(0,-1)(2.5,3)
\psset{arrowsize=0.2 2}
\psset{unit=1.1}
%
%
\pscoil[coilaspect=0,coilarm=0,coilwidth=0.25,coilheight=1.3,linecolor=black](0,0.3)(2.99,0)
\pscoil[coilaspect=0,coilarm=0,coilwidth=0.25,coilheight=1.3,linecolor=black](0,1.7)(2.99,2)
\put(1.5,2.2){$Z'$}
\put(1.5,-0.5){$W/Z$}
%
%
\psline[linecolor=black,linewidth=0.3mm]{->}(0,0.3)(0,1.2)
\psline[linecolor=black,linewidth=0.3mm]{-}(0,0.9)(0,1.7)
\put(-0.45,1){$q$}
%
%
\psline[linecolor=black,linewidth=0.3mm]{->}(-2,-0.5)(-0.9,-0.05)
\psline[linecolor=black,linewidth=0.3mm]{-}(-1,-0.1)(0,0.3)
\put(-1,2.3){$\bar{q}$}
\psline[linecolor=black,linewidth=0.3mm]{->}(0,1.7)(-1.1,2.105)
\psline[linecolor=black,linewidth=0.3mm]{-}(-1,2.08)(-2,2.5)
\put(-1,-0.5){$q$}
%
%
\psline[linecolor=black,linewidth=0.3mm](3,2)(3.75,2.37)
\psline[linecolor=black,linewidth=0.3mm]{<-}(3.4,2.2)(4,2.5)
\put(3.5,2.6){$\bar{\zeta}$}
\psline[linecolor=black,linewidth=0.3mm]{->}(3,2)(3.75,1.63)
\psline[linecolor=black,linewidth=0.3mm](3.5,1.75)(4,1.5)
\put(3.5,1.3){$\zeta$}
%
%
\psline[linecolor=black,linewidth=0.3mm]{->}(3,0)(3.75,0.37)
\psline[linecolor=black,linewidth=0.3mm](3.4,0.2)(4,0.5)
\put(3.5,0.55){$q$}
\psline[linecolor=black,linewidth=0.3mm](3,0)(3.75,-0.37)
\psline[linecolor=black,linewidth=0.3mm]{<-}(3.4,-0.2)(4,-0.5)
\put(3.5,-0.8){$\bar{q}$}
%
%
%
%
%
%
%
%
%
%
%
\end{pspicture}
%
%
\caption{\scshape{The $Z'$-decay into the pair $\bar{\zeta}-\zeta$.
The cascade effect as a possible Dark Matter detection via $W$- or $Z$-monojets.}} \label{Z'Decay}
\end{center}
\end{figure}
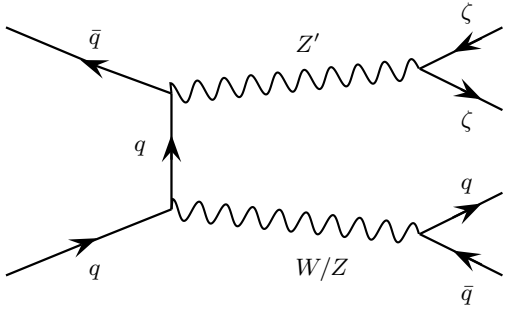
%


\noindent
Therefore, the previous rules and the QFT technical, the decay width of
$Z' \rightarrow \bar{\zeta} \, \zeta$ is given by
%
\begin{eqnarray}
\Gamma(Z' \rightarrow \bar{\zeta} \, \zeta)=\frac{e^2 \, m_{Z'}}{24\pi\sin^{2}(2\alpha) \cos^2\theta_{W}} \times
\nonumber \\
\times \, \sqrt{1-\frac{4m_{\zeta}^{\, 2}}{m_{Z'}^{\, 2}}} \left( 1
- \frac{3m_{\zeta}^{\, 2}}{4m_{Z'}^{\, 2}} \right) \, ,
\;\;\;
\end{eqnarray}
where $m_{Z'}>2 m_{\zeta}$. Using the previous values $m_{Z'}= 2 \, \mbox{TeV}$ and $m_{\zeta}=550 \, \mbox{GeV}$,
the $Z'$-width decay rate is
\begin{eqnarray}
\Gamma(Z' \rightarrow \bar{\zeta} \, \zeta) \simeq \frac{0.018}{\sin^2(2\alpha)} \, \mbox{TeV} \; .
\end{eqnarray}
This decay width is plotted below :
\begin{figure}[h]
\centering
\includegraphics[scale=0.44]{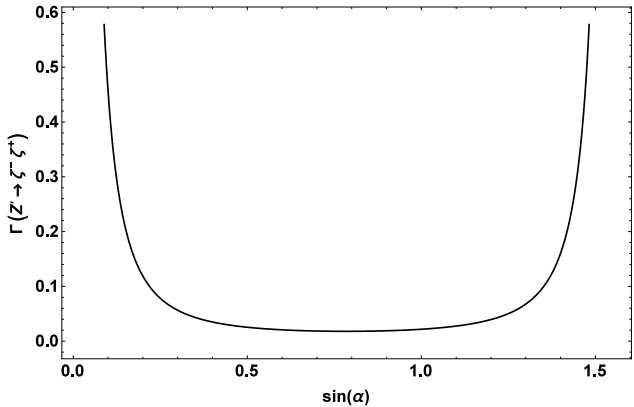}
\caption{The decay width of the process $Z' \rightarrow \bar{\zeta} \, \zeta$ plotted as function of the $\alpha$-mixing angle.}
\label{GraficoRelDispDFR}
\end{figure}

For the case of $\alpha=45^{o}$, the decay width is
$\Gamma(Z' \rightarrow \bar{\zeta} \, \zeta) \simeq 0.018 \, \mbox{TeV}$, and the $Z'$-decay time in this process is estimated by
\footnote{We have used the conversion formula $1 \, \mbox{TeV}=1.52 \times 10^{27} \, \mbox{s}^{-1}$ in the natural units $\hbar=c=1$. }
\begin{eqnarray}
\tau(Z' \rightarrow \bar{\zeta} \, \zeta)=\frac{1}{\Gamma(Z' \rightarrow \bar{\zeta} \, \zeta)}\simeq 3.7 \times 10^{-26} \, \mbox{s} \; .
\end{eqnarray}
%

\section{The $\mbox{MeV}$-scale physics and the $X$-boson}

In contrast to the previous case, we analyze the SSBs whenever $u \ll v$.
This sets a SSB at a lower scale $u$ with respect to $v = 246 \, \mbox{GeV}$ of the EW model.
Therefore, we introduce here the scenario in that the $VEV$ $u$-scale describes the
$\mbox{MeV}$-scale for $X$-boson of $m_{X}=17 \, \mbox{MeV}$, or the lightest para-photon at the $\mbox{Sub-eV}$ scale.
To accomplish this purpose, we re-start from the original symmetry $SU_{L}(2) \times U_{R}(1)_{J} \times U(1)_{K}$,
such that the sector of fermions is given by the Lagrangian (\ref{Lleptons}), but by convenience, the covariant derivatives
must exhibit the coupling $g^{\prime\prime}$ multiplied by a $\varepsilon$-parameter of weak magnitude to provide
the interaction of $C^{\mu}$ of $U(1)_{K}$ with the fermions of the Standard Model. It may connect the
Standard Model physics with a dark matter sector, see \cite{KitaharaPRD2017}. Thus, we modify
these couplings by introducing the real parameters $\varepsilon_{\Psi}$ and $\varepsilon_{\zeta}$ in connection with the gauge
coupling constant associated with the extra Abelian group, $U(1)_{K}$ :
\begin{eqnarray}\label{DmuPsichi}
D_{\mu} \Psi_{L} \!\!\! &=& \!\!\! \left(\partial_{\mu} \!+ \!i \, g \, A_{\mu}^{\, a} \, \frac{\sigma^{a}}{2} \!+\!i \, J_{L} \, g^{\prime} \, B_{\mu}
\!+\! i \, K_{L} \, \varepsilon_{\Psi} \, g^{\prime \prime} \, C_{\mu} \phantom{\frac{1}{2}} \!\!\!\!\! \right)\! \Psi_{L} \; ,
\nonumber \\
D_{\mu} \Psi_{R} \!\! &=& \!\! \left( \phantom{\frac{1}{2}} \!\!\!\! \partial_{\mu} + i \, J_{R} \, g^{\prime} \, B_{\mu}+i \, K_{R} \, \varepsilon_{\Psi} \, g^{\prime \prime} \, C_{\mu} \! \right) \Psi_{R} \; ,
\nonumber \\
D_{\mu} \nu_{R} \!\! &=& \!\! \left( \phantom{\frac{1}{2}} \!\!\!\! \partial_{\mu} + i \, J_{\nu_{R}} \, g^{\prime} \, B_{\mu}+i \, K_{\nu_{R}} \, \varepsilon_{\nu} \, g^{\prime \prime} \, C_{\mu} \! \right) \nu_{R} \; ,
\nonumber \\
D_{\mu}\zeta\!\! &=& \!\! \left( \phantom{\frac{1}{2}} \!\!\!\! \partial_{\mu} + i \, J_{\zeta} \, g^{\prime} \, B_{\mu} + i \, K_{\zeta} \, \varepsilon_{\zeta} \, g^{\prime \prime} \, C_{\mu} \right) \zeta \, ,
\end{eqnarray}
In the gauge sector, the extra-sector $U(1)_{K}$ of $C^{\mu}$ couples kinetically with $B^{\mu}$ of $U_{R}(1)_{J}$
by means of mixing $\chi$-parameter given by
\begin{equation}\label{Lgaugechi}
{\cal L}_{gauge}=-\frac{1}{2} \,\mbox{tr}\left(F_{\mu\nu}^{\; 2}\right)
-\frac{1}{4} \, B_{\mu\nu}^{\; 2}
-\frac{1}{4} \, C_{\mu\nu}^{\; 2}
+\frac{\chi}{2} \, B_{\mu\nu} \, C^{\mu\nu} \; .
\end{equation}
%
Its currently estimated value is $10^{-6} < \chi < 10^{-3}$ for models that discuss hidden photons as dark matter candidates \cite{Arias2010}.
In (\ref{Lgauge}), the mixing term $\chi \, B_{\mu\nu}C^{\mu\nu}$ was not considered because in that scenario, characterized high-energy effects,
we do not expect a tree-level mixing, as in the present scenario, where we intend to describe the photon-$X$-boson, or photon-para-photon mixing.
The Higgs sector of EW model looks like it was
proposed in (\ref{LHiggs}), but the extra-Higgs $\Xi$
is coupled to the $U(1)_{K}$ gauge boson through the $\varepsilon_{\Xi} \, g^{\prime\prime}$ coupling :
\begin{eqnarray}\label{DmuXiProto}
D_{\mu} \Xi =\left( \phantom{\frac{1}{2}} \!\!\!\!\! \partial_{\mu}+i \, J_{\Xi} \, g^{\prime} \, B_{\mu}
+i \, K_{\Xi} \, \varepsilon_{\Xi} \, g^{\prime\prime} \, C_{\mu} \! \right) \Xi \; .
\end{eqnarray}
After this SSB, we obtain the final symmetry breaking pattern :
\begin{equation}
SU_{L}(2) \times U_{R}(1)_{J} \times U(1)_{K} \stackrel{v}{\longmapsto}
U(1)_{G} \times U(1)_{K} \stackrel{u}{\longmapsto} U(1)_{em} \; ,
\end{equation}
where the group $U(1)_{G}$ is formed as a mixing of subgroups of $SU_{L}(2) \times U_{R}(1)_{J}$.
Notice the difference with respect to the previous case, where $U_{R}(1)_{J}\times U(1)_{K}$
was firstly broken. The free gauge sector, after this SSB, is represented by the Lagrangian
\begin{eqnarray}\label{LgaugeU1XU1BAmu3}
{\cal L}_{gauge} \!\!&=&\!\! -\frac{1}{2} \, W_{\mu\nu}^{+}W^{\mu\nu -}+ m_{W}^{\, 2} \, W_{\mu}^{+}W^{\mu -}
\nonumber \\
&&
\hspace{-1cm}
- \frac{1}{4} \, \left(\partial_{\mu}A_{\nu}^{\, 3}-\partial_{\nu}A_{\mu}^{\, 3} \right)^{2}
-\frac{1}{4} \, B_{\mu\nu}^{\,2}-\frac{1}{4} \, C_{\mu\nu}^{\,2}
\nonumber \\
&&
\hspace{-1cm}
+\frac{\chi}{2} \, B_{\mu\nu} \, C^{\mu\nu}
+\frac{v^{2}}{2} \left( J_{\Phi} \, g^{\prime} \, B_{\mu} - \frac{1}{2} \, g \, A_{\mu}^{\, 3} \right)^{2} .
\end{eqnarray}
Now, to render diagonal the $v$-mixed term in (\ref{LgaugeU1XU1BAmu3}), we introduce the transformations
\begin{eqnarray}\label{transfAYZ}
A_{\mu}^{\, 3} \!&=&\! \cos\theta_{W} \, \tilde{\tilde{Z}}_{\mu} + \sin\theta_{W} \, G_{\mu}
\nonumber \\
B_{\mu} \!&=&\! -\sin\theta_{W} \, \tilde{\tilde{Z}}_{\mu} + \cos\theta_{W} \, G_{\mu} \; ,
\end{eqnarray}
where $\theta_{W}$ satisfies the relation
%
$g_{G}=g \, \sin\theta_{W}=g^{\prime} \, \cos\theta_{W}$,
%
and $g_{G}$ is the coupling constant of $U(1)_{G}$-group. The neutral gauge sector of the model after this
diagonalization reads below :
\begin{eqnarray}\label{LgaugeU1XU1}
{\cal L}_{gauge} \!\!&=&\!\!
- \frac{1}{4} \, \tilde{\tilde{Z}}_{\mu\nu}^{\,2}
-\frac{1}{4} \, G_{\mu\nu}^{\,2}-\frac{1}{4} \, C_{\mu\nu}^{\,2}
-\frac{\chi}{2} \, \sin\theta_{W} \, \tilde{\tilde{Z}}_{\mu\nu} \, C^{\mu\nu}
\nonumber \\
&&
\hspace{-0.5cm}
+\frac{\chi}{2} \, \cos\theta_{W} \, G_{\mu\nu} \, C^{\mu\nu}
+\frac{1}{2} \,  m_{\tilde{\tilde{Z}}}^{\,2} \, \tilde{\tilde{Z}}_{\mu}^{\, 2} \; .
\end{eqnarray}
Using this diagonalization, the interactions of fermions with the neutral gauge bosons are given by
\begin{eqnarray}\label{LintGC}
{\cal L}^{int} \!\!&=&\!\! -\bar{\Psi}\left[\hspace{-0.3cm} \phantom{\frac{1}{2}} \left( g \, \cos\theta_{W} \, I^{3} \!-g' \, \sin\theta_{W} \, J \right) \slash{\!\!\!\!\tilde{\tilde{Z}}}
\right.
\nonumber \\
&&
\left.
\, + \, g_{G} \, \, G \, \, \, \slash{\!\!\!\!G}+ g'' \, \varepsilon_{\Psi} \, K \, \, \slash{\!\!\!\!C} \hspace{-0.3cm} \phantom{\frac{1}{2}} \, \right] \Psi \; .
\end{eqnarray}
The parametrization that defines $g_{G}$ suggests the definition for the generator of $U(1)_{G}$ as the sum $G=I^{3}+J$.
In this stage, we have two massless gauge fields $G^{\mu}$ and $C^{\mu}$, that is, we reach a gauge sector with gauge symmetry
$U(1)_{G}\times U(1)_{K}$.

To obtain a mass for the $X$-boson, we introduce the second $\Xi$-Higgs sector, with the covariant derivative (\ref{DmuXiProto}).
After this SSB with $u$-VEV scale, we get the gauge sector
\begin{eqnarray}\label{LgaugeU1XU1}
{\cal L}_{gauge} \!\!&=&\!\!
- \frac{1}{4} \, \tilde{\tilde{Z}}_{\mu\nu}^{\,2}
-\frac{1}{4} \, G_{\mu\nu}^{\,2}-\frac{1}{4} \, C_{\mu\nu}^{\,2}
-\frac{\chi}{2} \, \sin\theta_{W} \tilde{\tilde{Z}}_{\mu\nu} \, C^{\mu\nu}
\nonumber \\
&&
\hspace{-0.5cm}
+\frac{\chi}{2} \, \cos\theta_{W} \, G_{\mu\nu} \, C^{\mu\nu}
+\frac{1}{2} \,  m_{\tilde{\tilde{Z}}}^{\,2} \, \tilde{\tilde{Z}}_{\mu}^{\, 2}
\nonumber \\
&&
\hspace{-1.7cm}
+ \frac{u^2}{2} \! \left( \phantom{\frac{1}{2}} \hspace{-0.42cm}
-G_{\Xi} \, g^{\prime} \sin\theta_{W} \, \tilde{\tilde{Z}}_{\mu}
\!+\! G_{\Xi} \, g_{D} \, G_{\mu}
\!+\! K_{\Xi} \, \varepsilon_{\Xi} \, g^{\prime \prime} C_{\mu} \phantom{\frac{1}{2}} \!\!\!\!\!\!  \right)^2 \! ,
\hspace{0.5cm}
\end{eqnarray}
where we have used that $I^{3}=0$, and $J_{\Xi}=G_{\Xi}$ for the singlet $\Xi$-Higgs. The $G-C$ mixing suggests us to introduce
the following fields shift
\begin{eqnarray}
G_{\mu} \!\!&=&\!\!  A_{\mu}+\frac{\chi \cos\theta_{W} \tilde{\tilde{X}}_{\mu} }{\sqrt{1-\chi^{2} \cos^{2}\theta_{W}}}
\nonumber \\
C_{\mu} \!\!&=&\!\! \frac{\tilde{\tilde{X}}_{\mu}}{\sqrt{1-\chi^{2} \cos^{2}\theta_{W}}} \; ,
\end{eqnarray}
it cancel out the $G-C$ mixing term in the gauge sector to read the Lagrangian
\begin{eqnarray}\label{LgaugemassZAC}
{\cal L}_{gauge}\!\!&=&\!\!
-\frac{1}{4} \, F_{\mu\nu}^{\,2}
-\frac{1}{4} \, \tilde{\tilde{Z}}_{\mu\nu}^{\, 2}
-\frac{1}{4} \, \tilde{\tilde{X}}_{\mu\nu}^{\,2}
\nonumber \\
&&
\hspace{-1cm}
-\frac{\chi_{W}}{2} \, \tilde{\tilde{Z}}_{\mu\nu} \, \tilde{\tilde{X}}^{\mu\nu}
+\frac{1}{2} \, m_{Z}^{\, 2} \, \tilde{\tilde{Z}}_{\mu}^{\, 2}
+\frac{1}{2} \, m_{X}^{\, 2} \, \tilde{\tilde{X}}_{\mu}^{\, 2} \; ,
\end{eqnarray}
where the $\chi_{W}$-parameter is defined by
\begin{eqnarray}
\chi_{W} := \frac{\chi\, \sin\theta_{W}}{\sqrt{1-\chi^{2}\cos^{2}\theta_{W}}} \; ,
\end{eqnarray}
and the mass $m_{X}$ of $\tilde{\tilde{X}}^{\mu}$ is
%
$m_{X}= |K_{\Xi}| \, |\varepsilon_{\Xi}| \, e \, u $.
%
Here, it is also important to remember that we define the field strength tensors $\tilde{\tilde{Z}}^{\mu\nu}=\partial^{\mu}\tilde{\tilde{Z}}^{\nu}-\partial^{\nu}\tilde{\tilde{Z}}^{\mu}$,
$\tilde{\tilde{X}}^{\mu\nu}=\partial^{\mu}\tilde{\tilde{X}}^{\nu}-\partial^{\nu}\tilde{\tilde{X}}^{\mu}$ and $F^{\mu\nu}=\partial^{\mu}A^{\nu}-\partial^{\nu}A^{\mu}$.
The sector of interaction in (\ref{LintGC}) is written in terms of the gauge fields $A^{\mu}$ and $\tilde{\tilde{X}}^{\mu}$ as follows :
\begin{equation}\label{LintAZXtil}
{\cal L}^{int}=
- e \, Q_{em} \, \bar{\Psi} \; \slash{\!\!\!\! A} \, \Psi
- e \, Q_{\tilde{\tilde{Z}}} \, \bar{\Psi} \; \slash{\!\!\!\!\tilde{\tilde{Z}}} \, \Psi
-e \, Q_{X} \, \bar{\Psi} \; \slash{\!\!\!\!\tilde{\tilde{X}}} \, \Psi \; .
\end{equation}
The electric charge is so defined as : $Q_{em}=G=I^{3}+J$, and the fundamental charge obeys the parametrization
%
$e=g_{G}=g^{\prime\prime}$.
%
The $Q_{\tilde{\tilde{Z}}}$-charge generator of $Z$-boson like in (\ref{QZQZ'}), and $Q_{X}$ is given by
\begin{equation}\label{QX}
Q_{X}= +Q_{em} \, \chi \, \cos\theta_{W}+K_{\Psi} \, \varepsilon_{\Psi} \; ,
\end{equation}
where we have neglected terms of order $\chi^3$.
For convenience, the new $\Xi$-Higgs carries the charges $J_{\Xi}=0$ and $K_{\Xi}=+3$ to ensure a light $\zeta$-fermion
and also to avoid stable charged matter; see \cite{Duerr2014,Duerr2015}.
Here, the hypercharge $Y$ is not given by the sum of the charges $J$ and $K$.
It is so defined by the proper $J$-generator, {\it i. e.}, $J=Y$, and the $K$-generator has independent
values of the $Y$-charges. The simplest charge values are displayed in the \ref{Table2},
and the model in the X-boson scenario is also anomaly-free.
The $\zeta$-fermion does not carry electric charge, such that it just makes the $K$-charge
an example of hidden charge. Thus, this fact confirms the $\zeta$-fermion as a viable candidate to Dark Matter.
Other important point is that gauge symmetry forbids the mixed Yukawa interactions in (\ref{LHiggs}),
so we must take off the coupling constants, $z_{\nu_{\ell}}=t_{\nu_{\ell}} \rightarrow 0$. Therefore,
the gauge symmetry and the anomaly cancellation condition control the assignments of $K_{\zeta}$-charges :
$K_{\zeta_{L}}=-K_{\zeta_{R}}=+3/2$. At this stage, the model for the $X$-boson description proposed here
reduces to the analogous case of the $U_{B}(1)$-symmetry in the gauge sector, also studied in \cite{JFeng2016PRD}.
The charges are displayed in the table \ref{Table2}.
\begin{table}
\begin{tabular}{|l|l|l|l|l|l|l|}
\hline
\mbox{Fields} & $Q_{em}$ & $I^{3}$ & $Y=J$ & $K$ \\
\hline
\mbox{leptons-left} & $-1$ & $-1/2$ & $-1/2$ & $-1/2$  \\
\hline
\mbox{leptons-right} & $-1$ & $0$ & $-1$ & $-1$ \\
\hline
\mbox{neutrinos-left} & $0$ & $+1/2$ & $-1/2$ & $-1/2$ \\
\hline
\mbox{neutrinos-right} & $0$ & $0$ & $0$ & $0$ \\
\hline
$\zeta$-\mbox{fermion left} & $0$ & $0$ & $0$ & $+3/2$ \\
\hline
$\zeta$-\mbox{fermion right} & $0$ & $0$ & $0$ & $-3/2$ \\
\hline
\mbox{u-quark-left} & $+2/3$ & $+1/2$ & $+1/6$ & $+1/6$  \\
\hline
\mbox{d-quark-left} & $-1/3$ & $-1/2$ & $+1/6$ & $+1/6$ \\
\hline
\mbox{s-quark-left} & $-1/3$ & $-1/2$ & $+1/6$ & $+1/6$  \\
\hline
\mbox{u-quark-right} & $+2/3$ & $0$ & $+2/3$ & $+2/3$ \\
\hline
\mbox{d-quark-right} & $-1/3$ & $0$ & $-1/3$ & $-1/3$ \\
\hline
\mbox{s-quark-right} & $-1/3$ & $0$ & $-1/3$ & $-1/3$  \\
\hline
$W^{\pm}$-\mbox{bosons} & $\pm \, 1$ & $\pm \, 1$ & $0$ & $0$ \\
\hline
\mbox{neutral bosons} & $0$ & $0$ & $0$ & $0$  \\
\hline
$\Xi$-\mbox{Higgs} & $0$ & $0$ & $0$ & $+3$ \\
\hline
$\Phi$-\mbox{Higgs} & $0$ & $-1/2$ & $+1/2$ & $0$ \\
\hline
\end{tabular}
\caption{One solution anomaly-free for the particle content that can explains the $X$-boson scenario at the MeV-scale.}\label{Table2}
\end{table}
The massless gauge particle remaining in the model is associated to the $A^{\mu}$-field, so that we identify it as the EM photon.
It is not difficult to check that the gauge sector of (\ref{LgaugemassZAC}) is invariant under the $U(1)_{em}$ associated with the
gauge symmetry of $A^{\mu}$. The $\tilde{\tilde{Z}}$- and $\tilde{\tilde{X}}$-fields are not the physical fields corresponding to the
$Z$- and $X$-bosons yet. We carry out a diagonalization procedure to go over into the basis of physical fields.
The Lagrangian (\ref{LgaugemassZAC}) can be re-written into the matrix form as follows
\begin{eqnarray}\label{LgaugeMatriz}
{\cal L}_{\tilde{\tilde{Z}}-\tilde{\tilde{X}}}=\frac{1}{2} \left(\tilde{\tilde{V}}^{\mu}\right)^{t} \! \Box \theta_{\mu\nu} K \tilde{\tilde{V}}^{\nu}
+ \frac{1}{2} \left(\tilde{\tilde{V}}^{\mu}\right)^{\!t}\! \eta_{\mu\nu} M^{2} \tilde{\tilde{V}}^{\nu} ,
\end{eqnarray}
where $(\tilde{\tilde{V}}^{\mu})^{t}=\left( \; \tilde{\tilde{Z}}^{\mu} \; \; \tilde{\tilde{X}}^{\mu} \; \right)$, $K$ is the matrix
\begin{eqnarray}
K:=\left(
\begin{array}{cc}
1 & +\chi_{W}
\\
\\
+ \chi_{W} & 1 \\
\end{array}
\right)
\; .
\end{eqnarray}
The mass matrix $M^{2}$ is given by
\begin{eqnarray}
M^{2}=\left(
\begin{array}{cc}
m_{Z}^{\, 2} & 0
\\
\\
0 & m_{X}^{\, 2}
\end{array}
\right) \; .
\end{eqnarray}
To diagonalize the Lagrangian (\ref{LgaugeMatriz}), we carry out
an orthogonal transformation $\tilde{\tilde{V}} \; \longmapsto \; \tilde{V}= R \, \tilde{\tilde{V}} $, where $R^{t}\, R={\uma}$.
Thus, if we define the diagonal matrix as $K_{D}=R \, K \, R^{t}$, the eigenvalues of $K_{D}$ are given by
$\lambda_{\pm}=1 \pm \chi_{W}$, so we obtain the diagonal matrix
\begin{eqnarray}
K_{D}=
\left(
\begin{array}{cc}
1 + \chi_{W} & 0
\\
\\
0 & 1 - \chi_{W} \\
\end{array}
\right) \; .
\end{eqnarray}
In so doing, the Lagrangian in terms of $\tilde{V}^{\mu}$ takes over the form
\begin{equation}\label{LgaugeMatrizVtilKD}
{\cal L}_{\tilde{Z}-\tilde{X}}= \frac{1}{2} \, \left(\tilde{V}^{\mu}\right)^{\, t} \Box \, \theta_{\mu\nu} \, K_{D} \, \tilde{V}^{\nu}
+ \frac{1}{2} \, \left(\tilde{V}^{\mu}\right)^{\, t} \eta_{\mu\nu} \, \tilde{M}^{2} \, \tilde{V}^{\nu} \; ,
\end{equation}
where $\tilde{M}^{2}=R \, M^{2} \, R^{t}$. It can be readily checked that the solution for $R$ is the following $SO(2)$-matrix
\begin{eqnarray}
R=\frac{1}{\sqrt{2}} \,
\left(
\begin{array}{cc}
1 & 1
\\
\\
-1 & 1 \\
\end{array}
\right) \; ,
\end{eqnarray}
so, the matrix mass $\tilde{M}^{2}$ is given by
\begin{eqnarray}
\tilde{M}^{2}=\frac{1}{2}
\left(
\begin{array}{cc}
m_{Z}^{\,2} + m_{X}^{2} & m_{X}^{\,2}- m_{Z}^{\,2}
\\
\\
m_{X}^{\,2}- m_{Z}^{\,2} & m_{Z}^{\,2} + m_{X}^{2} \\
\end{array}
\right) \; .
\end{eqnarray}
Now, we write the matrix $K_{D}$ as $K_{D}=\left(K_{D}^{1/2}\right)^{t} \left( K_{D}^{1/2} \right)$
to adsorb it into the kinetic term, redefining $\tilde{V} \, \longrightarrow \, K_{D}^{1/2} \, \tilde{V}$.
The solution for the matrix $K_{D}^{1/2}$ is
\begin{eqnarray}
K_{D}^{1/2}=
\left(
\begin{array}{cc}
\sqrt{1+\chi_{W}} & 0
\\
\\
0 & \sqrt{1-\chi_{W}}
\end{array}
\right) \; .
\end{eqnarray}
Thus, the Lagrangian (\ref{LgaugeMatrizVtilKD}) is
\begin{equation}\label{LgaugeMatrizVtilKD}
{\cal L}_{\tilde{Z}-\tilde{X}}= \frac{1}{2} \left(\tilde{V}^{\mu}\right)^{\, t} \Box \, \theta_{\mu\nu} \, \tilde{V}^{\nu}
+ \frac{1}{2} \left(\tilde{V}^{\mu}\right)^{\, t} \eta_{\mu\nu} \, M_{D}^{\, 2} \, \tilde{V}^{\nu} \, ,
\end{equation}
where the mass matrix is now $M_{D}^{\, 2}=\left(K_{D}^{1/2}\right)^{-1} \tilde{M}^{2} \left(K_{D}^{1/2}\right)^{-1}$, that is,
\begin{eqnarray}\label{MDMass}
M_{D}^{2}=
\frac{1}{2}
\left(
\begin{array}{cc}
\frac{m_{Z}^{\, 2} + m_{X}^{2}}{1+\chi_{W}} & -\frac{m_{Z}^{\,2}- m_{X}^{\,2}}{\sqrt{1-\chi_{W}^{2}}}
\\
\\
-\frac{m_{Z}^{\,2}- m_{X}^{\,2}}{\sqrt{1-\chi_{W}^{\, 2}}} & \frac{m_{Z}^{\, 2} + m_{X}^{2}}{1-\chi_{W}} \\
\end{array}
\right) \; .
\end{eqnarray}
Since $M_{D}^{\, 2}$ is also real and symmetric, it can be diagonalized by an orthogonal matrix, $S$; we define $V^{\mu}=S \, \tilde{V}^{\mu}$:
\begin{eqnarray}
\left(
\begin{array}{c}
Z^{\mu} \\
X^{\mu} \\
\end{array}
\right)
=\left(
\begin{array}{cc}
\cos\theta_{ZX} & \sin\theta_{ZX}
\\
\\
-\sin\theta_{ZX} & \cos\theta_{ZX} \\
\end{array}
\right)
\left(
\begin{array}{c}
\tilde{Z}^{\mu} \\
\tilde{X}^{\mu} \\
\end{array}
\right) \; .
\end{eqnarray}
We end up with a fully diagonal Lagrangian as given below :
\begin{equation}\label{Lag3}
\mathcal{L}_{Z-X} = \frac{1}{2} \left( V^{\mu} \right)^{t} \Box\theta_{\mu\nu} V^{\nu}
+ \frac{1}{2} \left(V^{\mu}\right)^{t} \eta_{\mu\nu} M_{diag}^2 V^{\nu} \; ,
\end{equation}
with $M_{diag}^{\, 2} = S \, M_{D}^{\, 2} \,  S^{t}$ is given by the eigenvalues $M_{\pm}^{\, 2}$ of the mass matrix (\ref{MDMass}),
that is,
\begin{equation}
m_{Z-X}^{(\pm) \, 2}=\frac{m_{Z}^{\, 2}+m_{X}^{\, 2}\pm\sqrt{\left(m_{Z}^{2}-m_{X}^{2} \right)^{2}+4 \, m_{Z}^{2} \, m_{X}^{2} \chi_{W}^{2}} }{2\left(1-\chi_{W}^{\, 2}\right)} \; .
\hspace{0.3cm}
\end{equation}
The $\theta_{ZX}$ mixing angle satisfies the relation
\begin{eqnarray}
\tan2\theta_{ZX}= \left(\frac{m_{Z}^2-m_{X}^2}{m_{Z}^2+m_{X}^2} \right) \frac{\sqrt{1-\chi_{W}^{2}}}{\chi_{W}} \; .
\end{eqnarray}
Since we fix the mass of $Z$ at the $\mbox{GeV}$-scale, and $m_{X}$ is at $\mbox{MeV}$-scale or $\mbox{Sub-eV}$ for para-photons, we can assume the approximation $m_{X}/m_{Z} \ll 1$, so the eigenvalues matrix are reduced to expressions
\begin{eqnarray}
M_{Z} \!\!&=&\!\! m_{Z-X}^{\, (+)} \simeq \frac{m_{Z}}{\sqrt{1-\chi_{W}^{\, 2}}} \left( 1 + \frac{1}{2} \, \frac{m_{X}^{\, 2}}{m_{Z}^{\, 2}} \, \chi_{W}^{\, 2} \right) \, ,
\nonumber \\
M_{X} \!\!&=&\!\! m_{Z-X}^{\, (-)} \simeq \frac{m_{X}}{\sqrt{1-\chi_{W}^{\, 2}}} \left( \, 1 - \frac{\chi_{W}^2}{2} \, \right) \; ,
\end{eqnarray}
and the $\theta_{ZX}$-angle is $\tan2\theta_{ZX}\simeq \csc\theta_{W} \, \sqrt{1-\chi^2}/\chi$. Here, the $Z$-mass acquires a correction coming
from the mixing parameter $\chi_{W}$, and from the ratio $m_{X}/m_{Z}$, while the eigenvalue associated with the mass of the
$X$-boson is the same as in the previous expression. The $S$-matrix corresponds to an $SO(2)$-transformation that depends
on the $\chi_{W}$-parameter
%
\begin{eqnarray}
S=\frac{1}{\sqrt{2}}\left(
\begin{array}{cc}
\sqrt{1+\chi_{W}} & \sqrt{1-\chi_{W}}
\\
\\
-\sqrt{1-\chi_{W}} & \sqrt{1+\chi_{W}}
\end{array}
\right) .
\end{eqnarray}
Therefore, the transformation from the basis $\left\{ \, \tilde{\tilde{Z}}^{\mu} \, , \, \tilde{\tilde{X}}^{\mu} \, \right\}$ to the basis of
the physical $Z$- and $X$-bosons $\left\{ \, Z^{\mu} \, , \, X^{\mu} \, \right\}$ is represented by the inverse
$\tilde{\tilde{V}}=R^{t} \, (K_{D}^{1/2})^{-1} \, S^{t} \, V$, which implies in the shift
\begin{eqnarray}\label{transfZX}
\tilde{\tilde{X}}^{\mu} \!&=&\! X^{\mu}+\frac{\chi_{W} \, Z^{\mu}}{\sqrt{1-\chi_{W}^{\, 2}}}
\nonumber \\
\tilde{\tilde{Z}}^{\mu} \!&=&\! \, \frac{Z^{\mu}}{\sqrt{1-\chi_{W}^{\, 2}}} \; .
\end{eqnarray}
Thus, the full diagonal Lagrangian for the gauge sector is
\begin{eqnarray}
{\cal L}_{gauge}\!\!&=&\!\!-\frac{1}{2} \, W_{\mu\nu}^{+}W^{\mu\nu -}+ m_{W}^{\, 2} \, W_{\mu}^{+}W^{\mu -}
\nonumber \\
&&
\hspace{-0.5cm}
-\frac{1}{4} \, F_{\mu\nu}^{\,2}
-\frac{1}{4} \, Z_{\mu\nu}^{\, 2}+\frac{1}{2} \, M_{Z}^{\, 2} \, Z_{\mu}^{\, 2}
\nonumber \\
&&
\hspace{-0.5cm}
-\frac{1}{4} \, X_{\mu\nu}^{\,2}+\frac{1}{2} \, M_{X}^{\, 2} \, X_{\mu}^{\, 2} \, .
\end{eqnarray}
%


To estimate the kinetic mixing $\chi$-parameter and the $\varepsilon$-parameters, we analyze
the interactions between the $X$-boson and any chiral fermion of the model; this yields :
\begin{eqnarray}
{\cal L}^{\, int}=- \, e\, Q_{em} \, \bar{\Psi} \, \, \slash{\!\!\!\!A} \, \Psi
-e\, Q_{Z} \, \bar{\Psi} \, \, \slash{\!\!\!\!Z} \, \Psi
- e \, Q_{X} \, \bar{\Psi} \,\, \slash{\!\!\!\!X} \, \Psi \, .
\hspace{0.3cm}
\end{eqnarray}
Here, the millicharged generator $Q_{X}$ is like in (\ref{QX}), while the $Q_{Z}$-charge has a $\chi$-correction as follows
\begin{eqnarray}
Q_{Z}:= Q_{\tilde{\tilde{Z}}} + \chi \, \sin\theta_{W} \, Q_{X} \; ,
\end{eqnarray}
where we are neglecting terms of order-$\chi^3$. The $\Psi$-field is any fermion (quark or lepton) of the SM; it can also be the exotic $\zeta$-fermion.
Since $\Psi$ includes both the Left- and Right-components, we write the $X$-interaction in terms of vector and axial currents as given below :
\begin{equation}\label{LintX}
{\cal L}^{int}_{X}=-\, e \, \sum_{\Psi}\bar{\Psi} \, \left( \, c^{\Psi}_{V}+c^{\Psi}_{A} \, \gamma_{5} \, \right) \, \slash{\!\!\!\!X} \, \, \Psi \; ,
\end{equation}
where we define $c^{\Psi}_{V}$ and $c^{\Psi}_{A}$ as
\begin{eqnarray}
c^{\Psi}_{V}
\!\!&=&\!\! +Q_{em} \, \chi \, \cos\theta_{W}+\frac{1}{2}\left( K_{\Psi_{L}} \, \varepsilon_{\Psi_{L}}+K_{\Psi_{R}} \, \varepsilon_{\Psi_{R}} \right)
\nonumber \\
c^{\Psi}_{A} \!\!&=&\!\! \frac{1}{2} \left( K_{\Psi_{L}} \, \varepsilon_{\Psi_{L}}-K_{\Psi_{R}} \, \varepsilon_{\Psi_{R}} \right) \; .
\end{eqnarray}
The summation over $\Psi$ in (\ref{LintX}) runs over all fermions of the model.
As a consequence, we observe that the axial coefficients depend only on the $\varepsilon_{\Psi}$-parameters.
Using the charges cast in table (\ref{Table2}), we express the $c^{\Psi}_{V}$ and $c^{\Psi}_{A}$
as follows
\begin{eqnarray}
c^{\ell}_{V}\!&=&\! - \, \chi \, \cos\theta_{W}-\frac{1}{2}\left(\frac{1}{2} \, \varepsilon_{\ell_{L}}+ \varepsilon_{\ell_{R}} \right) \; ,
\nonumber \\
c^{\ell}_{A} \!&=&\! +\frac{1}{2}\left( -\frac{1}{2} \, \varepsilon_{\ell_{L}}+ \varepsilon_{\ell_{R}} \right) \; ,
\nonumber \\
c^{\nu}_{V} \!&=&\! c^{\nu}_{A}= -\frac{1}{4} \, \varepsilon_{\nu_{L}}
\hspace{0.2cm} , \hspace{0.2cm}
\nonumber \\
c^{u}_{V}\!\!&=&\!\! +\frac{2}{3} \, \chi \, \cos\theta_{W}
+\frac{1}{2}\left( \frac{1}{6} \, \varepsilon_{u_{L}}+\frac{2}{3} \, \varepsilon_{u_{R}} \right)
\hspace{0.2cm} , \hspace{0.2cm}
\nonumber \\
c^{u}_{A} \!\!&=&\!\! +\frac{1}{2}\left( \frac{1}{6} \, \varepsilon_{u_{L}}-\frac{2}{3} \, \varepsilon_{u_{R}} \right)
\nonumber \\
c^{d}_{V}\!\!&=&\!\! -\frac{1}{3} \, \chi \, \cos\theta_{W}+\frac{1}{2}\left( \frac{1}{6} \, \varepsilon_{d_{L}}-\frac{1}{3} \, \varepsilon_{d_{R}} \right)
\hspace{0.2cm} , \hspace{0.2cm}
\nonumber \\
c^{d}_{A} \!\!&=&\!\! +\frac{1}{2}\left( \frac{1}{6} \, \varepsilon_{d_{L}}+\frac{1}{3} \, \varepsilon_{d_{R}} \right)
\nonumber \\
c^{s}_{V}\!\!&=&\!\! -\frac{1}{3} \, \chi \, \cos\theta_{W}+\frac{1}{2}\left( \frac{1}{6} \, \varepsilon_{s_{L}}-\frac{1}{3} \, \varepsilon_{s_{R}} \right)
\nonumber \\
c^{s}_{A} \!\!&=&\!\! +\frac{1}{2}\left( \frac{1}{6} \, \varepsilon_{s_{L}}+\frac{1}{3} \, \varepsilon_{s_{R}} \right)
\nonumber \\
c^{\zeta}_{V} \!\!&=&\!\! +\frac{3}{2} \, \left(\varepsilon_{\zeta_{L}}-\varepsilon_{\zeta_{R}}\right)
\hspace{0.1cm} , \hspace{0.1cm}
c^{\zeta}_{A}= +\frac{3}{2} \, \left(\varepsilon_{\zeta_{L}}+\varepsilon_{\zeta_{R}}\right) \, .
\end{eqnarray}
We then use quantum field-theoretic rules to work out the expression for the $X$-decay width into any $\Psi$-fermion:
\begin{eqnarray}
\Gamma(X \rightarrow \bar{\Psi} \, \Psi )&&\!\!\!\!\!=\frac{e^2 \, m_{X} }{24\pi}
\left( \, |c^{\Psi}_{V}|^2+|c^{\Psi}_{A}|^2 \, \right)
\times
\nonumber \\
&&
\hspace{-0.4cm}
\times \,
\sqrt{1-\frac{4m_{\Psi}^{\, 2}}{m_{X}^{\, 2}}} \left( 1
- \frac{3}{4}\frac{m_{\Psi}^{\, 2}}{m_{X}^{\, 2}} \right) \; ,
\end{eqnarray}
by virtue of the condition $m_{X} > 2 \, m_{\Psi}$.

We notice that the vector and axial coefficients depend on the $11$
$\varepsilon$-parameters due to the coupling of the fermions with the dark
sector $U(1)_{K}$-factor. Since we have vector and axial currents in (\ref{LintX}),
it is important to remember that in general, the protophobic property involving nucleons
interaction must not emerge here. Therefore, we can study conditions on $\varepsilon$-parameters in the sectors
of leptons and quarks. In the leptonic sector, the current is purely vector-like whenever
$\varepsilon_{\ell_{L}}=2 \, \varepsilon_{\ell_{R}}$, thus the $c_{V}^{\ell}$ coefficient
is $c_{V}^{\ell}=- \, \chi \, \cos\theta_{W}-\varepsilon_{\ell_{R}}$. On the other hand,
the condition for purely axial current is $-4 \, \chi \, \cos\theta_{W}=\varepsilon_{\ell_{L}}+ 2 \, \varepsilon_{\ell_{R}}$.
We consider the simplest case in which the $X$-boson has both axial- and vector-type couplings with leptons.
For this, we fix the $\varepsilon$-parameters as $\varepsilon_{\ell_{L}}=-2 \, \varepsilon_{\ell_{R}}$;
consequently, we obtain the coefficients $c_{V}^{\ell}=-\chi \, \cos\theta_{W}$ and $c_{A}^{\ell}=-\varepsilon_{\ell_{L}}/2$.
Under these conditions, the decay width for the process $X \, \rightarrow \, e^{+} \, e^{-}$
is worked out and it can read as follows below :
\begin{eqnarray}
\Gamma(X \, \rightarrow \, e^{+} \, e^{-})=0.02 \, \left( \chi^2 \, \cos^2\theta_{W}+ \frac{\varepsilon_{e_{L}}^{2}}{4} \right)
 \, \mbox{MeV} \, ,
\end{eqnarray}
in which the KLOE-2 collaboration \cite{KLOE2} sets a limit of
\begin{eqnarray}
\chi_{eff}^{e}=\sqrt{\chi^2 \, \cos^2\theta_{W}+ \frac{\varepsilon_{e_{L}}^{2}}{4}} \lesssim 2 \times 10^{-3} \; .
\end{eqnarray}
The M${\o}$ller scattering yields parity violation from the mixed axial-vector couplings of the $X$-boson to leptons.
There emerges the constraint \cite{KahnJHEP2017}
\begin{eqnarray}
| \, \chi \, \varepsilon_{e_{L}} \, | \lesssim 2.27 \times 10^{-8} \; .
\end{eqnarray}
Neutrinos interact with the $X$-boson through the $\varepsilon_{\nu_{L}}$-parameter. Both axial and vector couplings
disappear whenever $\varepsilon_{\nu_{L}} \rightarrow 0$. The $X$-boson decay into the $\nu_{e}$-neutrino is therefore expressed by the width
\begin{eqnarray}\label{Decaynu}
\Gamma(X \rightarrow \bar{\nu}_{e} \, \nu_{e} )
=\varepsilon_{\nu_{e}}^{2} \frac{e^2 \, m_{X} }{192 \pi}
= 2.5 \times 10^{-3} \, \varepsilon_{\nu_{e}}^{2} \, \mbox{MeV} \; ,
\end{eqnarray}
in which $m_{X} \gg m_{\nu}$. The most neutrino constraint is in the neutrino-electron scattering
$\bar{\nu}_{e} \, e \, \rightarrow \, \bar{\nu}_{e} \, e$. Thereby, the constraint fixes the bound \cite{BilmisPRD2015}
\begin{eqnarray}
\sqrt{| \, \varepsilon_{e} \, \varepsilon_{\nu_{e}} \, |} < 3 \times 10^{-4} \; .
\end{eqnarray}
The $X$-interaction with the exotic $\zeta$-fermion carries the $\varepsilon_{\zeta}$-parameters:
%
%
the corresponding decay width is
\begin{eqnarray}
\Gamma(X \, \rightarrow \, \bar{\zeta} \, \zeta ) \!&=&\! \frac{3 \, e^2}{16\pi} \, m_{X} \left( \varepsilon_{\zeta_{L}}^2+\varepsilon_{\zeta_{R}}^2 \right)
\nonumber \\
\!&=&\! 0.09 \, \left( \varepsilon_{\zeta_{L}}^2+\varepsilon_{\zeta_{R}}^2 \right) \, \mbox{MeV} \; .
\end{eqnarray}
In the sector of quarks, the $X$-boson has only axial couplings to avoid flavour mixing in the first generation. So,
under these conditions, the $\varepsilon$-parameters of the $u$- , $d$- and $s$-quarks are related as follows
\begin{eqnarray}\label{condaxial}
\varepsilon_{u_{L}}+2 \, \varepsilon_{d_{L}} \!\!&=&\!\! 4 \left( \varepsilon_{d_{R}}- \varepsilon_{u_{R}} \right)
\nonumber \\
\varepsilon_{u_{L}}+2 \, \varepsilon_{s_{L}} \!\!&=&\!\! 4 \left( \varepsilon_{s_{R}}- \varepsilon_{u_{R}} \right)
\nonumber \\
\varepsilon_{d_{L}}- \varepsilon_{s_{L}} \!\!&=&\!\! 2 \left( \varepsilon_{d_{R}}- \varepsilon_{s_{R}} \right) \; .
\end{eqnarray}
The $\chi_{p}$- and $\chi_{n}$-parameters for the proton and neutron are obtained using the nuclear matrix elements
associated with the correspondent operator from the axial current in (\ref{LintX}). Thus, the operator is putted between
the nucleon state $| N \rangle \, \left( \, N=n \, , \, p \, \right)$, such that
\begin{eqnarray}
\langle N | \, \sum_{Q} \, c_{A}^{Q} \, \bar{Q} \, \gamma_{5} \, \gamma^{\mu} \, Q \, | N \rangle=
\delta_{i}^{\,\, \mu} \, \sigma^{i} \, \sum_{Q} \, c_{A}^{Q} \, \Delta Q^{(N)} \; ,
\end{eqnarray}
where $Q$ runs over quarks flavours as $Q=\left\{ \, u \, , \, d \, , \, s \, \right\}$. This result is like
that one obtained in \cite{KozaczukPRD2017}. The $\Delta Q^{(N)}$ coefficients are numerically given by
\begin{eqnarray}
\Delta u^{(p)} \!&=&\! \Delta d^{(n)}=0.897 (27)
\nonumber \\
\Delta d^{(p)} \!&=&\! \Delta u^{(n)}=-0.367 (27)
\nonumber \\
\Delta s^{(p)} \!&=&\! \Delta s^{(n)}=-0.026 (4) \; .
\end{eqnarray}
Therefore, the axial current operator is written in the isospin notation
\begin{eqnarray}
{\cal L}^{int}_{X-N}=\frac{1}{2} \, \bar{N} \, \left( \vec{\sigma} \cdot \vec{X} \right) \left( \, \chi_{0} \, {\uma}+\chi_{1} \, \sigma^{3} \, \right) \, N \; ,
\end{eqnarray}
where $N=\left( \, p \; \; n \, \right)^{t}$, $\sigma^{3}$ is the Pauli matrix, and the parameters $\chi_{0}$ and $\chi_{1}$ are defined by
\begin{eqnarray}
\chi_{0} \!\!&=&\!\! \left(\Delta u^{(p)}+\Delta d^{\,(p)}\right)\left(c_{A}^{u}+c_{A}^{d}\right)+2 \, \Delta s^{(p)} \, c_{A}^{s}
\nonumber \\
\chi_{1} \!\!&=&\!\! \left(\Delta u^{(p)}-\Delta d^{\,(p)}\right)\left(c_{A}^{u}-c_{A}^{d}\right) \; .
\end{eqnarray}
Thereby, the $\chi$-coefficients for proton and neutron are $\chi_{p}=(\chi_{0}+\chi_{1})/2$ and $\chi_{n}=(\chi_{0}-\chi_{1})/2$, respectively,
so the results in terms of $c_{A}^{Q}$ are given by
\begin{eqnarray}
\chi_{p} \!\!&=&\!\! \Delta u^{(p)}c_{A}^{u}+\Delta d^{\,(p)}c_{A}^{d}+ \Delta s^{(p)} \, c_{A}^{s}=\sum_{Q} c_{A}^{Q} \, \Delta Q^{(p)} \,
\nonumber \\
\chi_{n} \!\!&=&\!\! \Delta u^{(p)}c_{A}^{d}+\Delta d^{\,(p)}c_{A}^{u}+ \Delta s^{(p)} \, c_{A}^{s} \; .
\end{eqnarray}
The $\varepsilon$-parameters are so estimated by using the allowed region for the quarks couplings required to explain the
$8 \, Be$-anomaly with a light axial vector, for more details, see \cite{KozaczukPRD2017}. Assuming the conditions
$c_{A}^{\, u}<0$, $c_{A}^{\, d}>0$ and $c_{A}^{\, d}=c_{A}^{\, s}$, the allowed region fixes the $\varepsilon$-parameters following the constraints below :
\begin{eqnarray}
&&
10^{-4} \lesssim |\varepsilon_{d_{L}}+2 \, \varepsilon_{d_{R}}| \lesssim 10^{-3}
\nonumber \\
&&
10^{-4} < |\varepsilon_{u_{L}}-4 \, \varepsilon_{u_{R}}| \lesssim 10^{-3} \; .
\end{eqnarray}
We can also fix up the $u$-scale by taking the mass of $m_{X}=17 \, \mbox{MeV}$ as the relation
$u \simeq 19 \, |\varepsilon_{\Xi}|^{-1} \, \mbox{MeV}$, since the $\varepsilon$-parameters are estimated in the range $10^{-4}-10^{-3}$,
the new VEV $u$-scale is bounded by $u= (1.9-19) \, \mbox{GeV}$.
The eigenvalues of (\ref{autovaloresMHF}), when $m_{\tilde{H}} \gg m_{\tilde{F}}$, yield the masses in the Higgs sector
\begin{eqnarray}
M_{H}=m_{F-H}^{\,(-)} \!\!&\simeq&\!\! \sqrt{ 2 \, \lambda_{\Phi} \, v^{2}}
\, \left( 1+\frac{\lambda^2}{8 \, \lambda_{\Phi}^2} \, \frac{u^2}{v^2} \right) \; ,
\nonumber \\
M_{F}=m_{F-H}^{\,(+)} \!\!&\simeq&\!\! \sqrt{ 2 \, \lambda_{\Xi} \, u^{2}}
\, \left( 1-\frac{\lambda^2}{8 \, \lambda_{\Phi} \, \lambda_{\Xi}} \right) \; ,
\end{eqnarray}
%
so that the mass of the $F$-Higgs is estimated in the range
%
$1.9 \, \mbox{GeV} \lesssim M_{F} \lesssim 19 \, \mbox{GeV}$.
%
In this ($X$-boson) scenario, since we have to impose the Yukawa couplings $t_{\nu_{\ell}}=z_{\nu_{\ell}}\rightarrow 0$ to ensure gauge invariance, the eigenvalues in (\ref{mlzeta}),
for $|g_{f}| \rightarrow 0$, are $m_{\zeta}= \frac{|w| \, u}{\sqrt{2}}$ and $m_{\nu_{\ell}}=\frac{|x_{\nu_\ell}| \, v}{\sqrt{2}}$.
Taking the $|w|$-coupling constant as in the $Z'$-scenario, {\it i. e.}, $|w| \simeq 0.27$, and considering the
estimation we have for the $u$-scale, we get the $\zeta$-fermion mass in the range $m_{\zeta}=0.36-3.6 \, \mbox{GeV}$, which respects the Tremaine-Gunn lower bound
\cite{DasPRD2012}, if we wish to interpret the $\zeta$-fermion as a particle present in the dark matter sector.
So, as we were already expecting, the $\zeta$-fermion in the $X$-boson scenario is lighter its counterpart in the $Z'$-scenario.




To conclude this Letter, we would like to summarize that, by adopting an anomaly-free $SU_{L}(2)\times U_{R}(1)_{J} \times U(1)_{K}$- type model,
with two Higgs fields, we may set up two different scenarios : in a case, we get the situation of a $\mbox{TeV}$-scale physics that describes
a hypothetic $Z'$-particle with a mass around the $2 \, \mbox{TeV}$. On the other hand,
changing the symmetry breaking pattern, our model may be describing a physical landscape in the $\mbox{MeV}$-scale of the recently proposed
$X$-boson associated with the $8\, \mbox{Be}^{\star}$-decay. In this context, the exotic $\zeta$-fermion must be a particle candidate to the
Dark Matter content at the Sub-eV scale. The phenomenology involving the interactions of the $X$-boson with neutrinos and the $\zeta$-fermion
is a problem to be discussed in a forthcoming paper.


%
%

%



%



\end{document}